\documentclass[11pt,a4paper]{article}
\pdfoutput=1
\usepackage{jheppub}
\usepackage{amsmath,amssymb,amsfonts}
\usepackage{enumitem,kantlipsum,mathdots}
\usepackage{floatrow}

\newcommand{\e}{\epsilon}

\renewcommand{\L}{\mathcal{L}}
\newcommand{\bcL}{\bar{\mathcal{L}}}

\newcommand{\h}{{\bar h}}

\newcommand{\p}{\partial}

\newcommand{\s}{\sigma}
\newcommand{\D}{\Delta}
\newcommand{\refb}[1]{(\ref{#1})}

\renewcommand{\>}{\rangle}

\def\beaa{\begin{eqnarray*}}
\def\eeaa{\end{eqnarray*}}
\def\bee{\begin{equation*}}
\def\eee{\end{equation*}}
\def\bea{\begin{eqnarray}}
\def\eea{\end{eqnarray}}
\def\be{\begin{equation}}
\def\ee{\end{equation}}
\def\ba{\begin{align}}
\def\ea{\end{align}}

\def\vec{\overrightarrow}

\title{BMS Characters and Modular Invariance}

\author[a,b]{Arjun Bagchi,} \author[a]{Amartya Saha,} 

\author[a]{and Zodinmawia.} \author{\\}

\affiliation[a]{Indian Institute of Technology Kanpur, Kalyanpur, Kanpur 208016. INDIA\\} 

\affiliation[b]{Centre for Particle Theory, Department of Mathematical Sciences,
Durham University, \\ South Road, Durham DH1 3LE, UK \\}

\emailAdd{abagchi@iitk.ac.in, amartyas@iitk.ac.in, zodin@iitk.ac.in}

\abstract{We construct the characters for the highest weight representations of the 3d Bondi-Metzner-Sachs (BMS$_3$) algebra. We reproduce our character formula by looking at singular limits from 2d CFT characters and find that our answers are identical to the characters obtained for the very different induced representations. We offer an algebraic explanation to this arising from an automorphism in the parent 2d CFT. We then use the characters to construct the partition function and show how to use BMS modular transformations to obtain a density of primary states. The entropy thus obtained accounts for the principal part of the entropy obtained from the BMS-Cardy formula. This suggests that BMS primaries capture most of the entropy of Flat Space Cosmologies, which are the flatspace analogues of BTZ black holes in AdS$_3$. } 

\begin{document}

\maketitle

\section{Introduction}
Conformal field theories (CFTs) \cite{DiFrancesco:1997nk} are a high energy theorist's dream. Symmetries of a relativistic conformal theory are constraining enough to determine many quantities of interest, e.g. the form of the two and three point functions in arbitrary dimensions, even without requiring the details of the underlying Lagrangian description. Powerful methods of the conformal bootstrap \cite{Ferrara:1973yt, Polyakov:1974gs, Rattazzi:2008pe} (for a modern introduction see e.g.~\cite{Simmons-Duffin:2016gjk}), which relies on the crossing symmetry of 4-point functions, constrain the system further and quite severely, and these are being currently utilised to chart out the allowed parameter space of all relativistic CFTs. The spectrum of primary operators and the coefficients of the three point functions (and the central charge) is all the information that is required to specify a CFT.  The bootstrap equation tells us which of these sets of data constitute CFT consistent with fundamental requirements like crossing symmetry. As is well known, this programme of constraining CFTs has far reaching consequences. The modern way of understanding all relativistically invariant quantum field theories (QFTs) is by renormalisation group flows away from fixed points governed by CFTs. So a classification of all CFTs in a sense would lead to the classification of all QFTs. 

\medskip

\noindent{\em{Conformal field theories and two dimensions}}

\medskip

\noindent
In two spacetime dimensions, CFTs become even more special \cite{Belavin:1984vu}. The underlying symmetry algebra is enhanced to two copies of the infinite dimensional Virasoro algebra. When defined on the 2d plane, methods of complex variables allow us to have tremendous analytic control on the theory and the 2d theory holomorphically factorises into a chiral and an anti-chiral sector, which are treated separately. There seems to be no prior restriction to how one should put  the chiral and anti-chiral sectors together. 2d CFTs arise in the context of string theory, as the residual symmetries on the string worldsheet after the fixing of conformal gauge, and the formulation on the complex plane, or equivalently the Riemann sphere, corresponds to tree-level string scattering. Loop diagrams are, of course, an integral part of any scattering computation and the higher loop contributions arise in string theory by placing 2d CFT on higher genus surfaces. Consistency on higher genus Riemann surfaces impose further conditions on 2d CFTs. For one loop diagrams, the surface of interest is the torus and these consistency requirements lead to modular invariance of 2d CFTs. Of late, these consistency requirements have led to what is called the modular bootstrap and has contributed in further constraining 2d CFTs. 

Modular invariance of 2d CFTs not only constrains possible CFT data, but has led to the famous Cardy formula which computes the entropy of the theory by relating the high energy regime to the low energy by specifically the modular S-transformation \cite{Cardy:1986ie}. In the context of holography, the Cardy formula has been used to great effect in matching up with the entropy of the BTZ black holes in the bulk dual AdS$_3$ theory \cite{Strominger:1997eq, Carlip:1998qw}. Recently, these ideas have been generalised to obtain averaged three-point function coefficients of two heavy and one light operator based on modular properties of 1-point torus amplitudes \cite{Kraus:2016nwo, Romero-Bermudez:2018dim, Hikida:2018khg, Brehm:2018ipf, Das:2017vej}. It has also been shown that crossing symmetry can be translated to modular properties, making this another avenue of implementing bootstrap-like techniques \cite{Maldacena:2015iua, Das:2017cnv}. 

\medskip

\noindent{\em{Holography and flatspace}}

\medskip

\noindent Holography in AdS spacetime suggests that the asymptotic symmetry group of the gravitation theory dictates the symmetries of the putative dual boundary theory. A natural way to extend the notion of holography to non-AdS spacetimes is thus the following: one should use canonical methods to calculate the asymptotic symmetry group of a bulk theory. This is the group of allowed diffeomorphisms given a particular set of boundary conditions, modded out by the trivial diffeos (the ones which lead to zero charge). One should then attempt to realise this as the symmetry group of a field theory that lives on the asymptotic boundary of that spacetime. 

In particular, we are interested in formulating the dual theory of asymptotically flat spacetimes. The asymptotic symmetry group of asymptotically Minkowski spacetime at its null boundary is given by the infinite dimensional Bondi-Metzner-Sachs (BMS) group \cite{Bondi:1962px, Sachs}. This flies in the face on conventional wisdom, which would have suggested that the ASG would be the Poincare group. It has been shown that even though in higher dimensions ($D>4$) using stringent boundary conditions one can restrict to the Poincare group, in $D=3, 4$, the infinite dimensional group is unavoidable. While taken to be mostly a curiosity after its initial discovery in the 1960's, it has been released in the recent past that the BMS symmetries have a fundamental role to play in the physics of the infra-red and link soft graviton theorems and memory effects in a triangle of relations (for a review of recent developments, see e.g. \cite{Strominger:2017zoo}).   

In this paper, we will be interested in BMS$_3$ \cite{Barnich:2006av} and its realisation in field theories in $D=2$, which according to the formulation above, will lead to a dual theory of 3d flat space, living on its null boundary \cite{Bagchi:2010zz} {\footnote{For higher dimensional putative dual theories to flat space, see discussions in e.g. \cite{Bagchi:2016bcd, Bagchi:2019xfx}.}}. Specifically, we would be interested in the construction of characters of the theory and the notion of modularity in these field theories. The calculation of characters help us construct the partition function of the BMS field theory. In a manner similar to 2d CFTs, the partition functions of 2d theories also admit a deformed version of modular invariance, which leads to a BMS-Cardy formula \cite{Bagchi:2012xr}. There are solutions in 3D flatspace which are obtained by quotienting Minkowski spacetime by a boost and a translation \cite{Cornalba:2003kd}. These are cosmological solutions called Flat Space Cosmologies, which have a cosmological horizon. The Bekenstein-Hawking entropy of these FSCs are reproduced by the BMS-Cardy formula \cite{Bagchi:2012xr, Barnich:2012xq}. 

\medskip

\noindent{\em{BMS, strings and other things}}

\medskip

\noindent
In connection with the earlier motivation of string theory, it is of interest to note that the same algebra arises as residual symmetries on the worldsheet of the tensionless bosonic string \cite{Isberg:1993av, Bagchi:2013bga,Bagchi:2015nca}. So this notion of modularity would be of interest when one considers scattering of strings in the tensionless limit. In a series of famous papers Gross and Mende \cite{Gross:1987kza,Gross:1987ar}, and later Gross \cite{Gross:1988ue} found that there was a very large simplification in the behaviour of string scattering amplitudes in this limit, leading to an infinite number of linear relations between scattering amplitudes of different string states valid order by order in perturbation theory. This pointed to some higher symmetry structure in this extreme stringy limit. Modular invariance of BMS would be central to understanding this Gross-Mende regime from the point of view of the symmetries on the worldsheet. 

\smallskip

\noindent Intriguingly, it has been recently found that these same BMS$_3$ symmetries arise in the formulation of the ambi-twistor string \cite{Casali:2016atr, Casali:2017zkz}. Ambitwistor strings \cite{Mason:2013sva} are a variant of the original twistor string theory that provide a basis for understanding of the Cachazo-He-Yuan (CHY) formula for scattering of massless fields in quantum field theories \cite{Cachazo:2013hca}. Modular invariance of BMS constructed earlier in \cite{Bagchi:2012xr, Bagchi:2013qva} would find its uses here as well, as has been noted in \cite{Casali:2017zkz}.  

\smallskip

\noindent Finally, BMS$_3$ algebras are isomorphic to 2d Galilean CFTs \cite{Bagchi:2010zz}. In a manner similar to relativistic CFTs, Galilean CFTs \cite{Bagchi:2009my, Bagchi:2009pe} would govern renormalization group flow fixed points for Galilean invariant quantum field theories. So the bootstrap programme we have initiated in \cite{Bagchi:2016geg}, augmented by constraints arising out of modularity, would help in charting out the allowed space of GCFTs and thereby Galilean field theories.

\medskip

\noindent {\em{A short summary of the paper}}

\medskip

\noindent In this paper, we formulate the construction of characters of the BMS$_3$ algebra in the highest weight representations. We show two different ways of arriving at the formula for the character in these highest weight representation, one of which follows from observations in the algebra and the other relies on the construction of the Gram matrix of inner products. As a robust cross-check of our answers, we reproduce the same characters in singular limits of 2d CFT answers. We then comment at length on the surprising relation of the characters in the highest weight representations of the BMS$_3$ with those of the very different induced representations computed earlier in \cite{Oblak:2015sea} and reproduced by a calculation of 1-loop determinants in 3d flat space \cite{Barnich:2015mui}. 
Through our analysis, we find that the character makes sense as a quantity that counts the number of possible states at a given level in the 2d field theory. This is a strong suggestion that the BMS-Cardy formula which accounts for the entropy of cosmological horizons in 3d flat spacetimes, could have a microscopic origin along the lines of the Strominger-Vafa construction \cite{Strominger:1996sh}. 

\medskip

\noindent {\em{Outline of the paper}}

\medskip
\noindent We start in Sec~2 with a brief review of the building blocks of the holographic correspondence in 3d asymptotically flat space. In Sec ~2.1, we discuss the properties of the putative 2d field theory invariant under the BMS$_3$ algebra. The representation of interest, the highest weight representation, is introduced and elaborated on. We also touch upon correlation functions, the construction of the partition function and the BMS-Cardy formula for the counting of states. In Sec~2.2, we move on to aspects of bulk 3d physics and introduce Flat Space Cosmologies (FSC) of asymptotically flat spacetime and comment on how the BMS-Cardy formula captures the entropy of the cosmological horizon of the FSCs. 

In Sec~3, we derive the character formula for these highest weight representations, first from the commutation relations of the underlying algebra alone and then by looking at the Gram matrix of inner products and constructing the trace of the operator from there. The Gram matrix for the BMS throws up some interesting structure, the details of which we delve into in the two appendices at the end of the paper. 

In Sec~4, we reproduce the character formula earlier obtained by intrinsically BMS methods, by singular limits from 2d CFT. We remind the reader of the two limits, the non-relativistic and the ultra-relativistic, that take one from the two copies of the Virasoro algebra to the BMS$_3$ algebra. The character formula is reproduced by both these limits. We point out why this is a big surprise and then go on to offer an explanation based on a novel automorphism in the parent 2d CFT. 

In Sec~5, as an application of the characters obtained, we derive a formula for the density of primary states, based on the deformed modular properties of the BMS-invariant field theory. We construct the partition function based on the characters of the highest weight representations and then impose modularity on this to obtain a BMS-Cardy formula for just the primaries of the field theory. In the holographic limit, this matches with the usual BMS-Cardy formula and thus we can say that the majority of the states contributing to the entropy of FSCs are BMS primaries. We conclude in Sec~6 with a summary of the results and a discussion of future directions of work.

\section{BMS$_3$ holography: a lightning review}
In this section, we briefly review some important aspects of Minkowskian holography in 3d bulk and 2d boundary case. 

\subsection{Aspects of dual 2d field theory}
The asymptotic symmetry group for Einstein gravity in 3d asymptotically flat spacetimes at its null boundary is given by the BMS group, the associated algebra of which is given by \cite{Barnich:2006av}
\bea
&& [L_n, L_m] = (n-m) L_{n+m} + {c_L}\delta_{n+m, 0} (n^3 - n) \crcr
&& [L_n, M_m] = (n-m) M_{n+m} + {c_M}\delta_{n+m, 0} (n^3 - n), \crcr
&& [M_n, M_m] = 0.
\label{bms3}
\eea
As motivated in the introduction, we would like to construct the notion of 3D flat holography by demanding that there exists a putative dual 2D field theory living on the null boundary that inherits this asymptotic BMS$_3$ algebra as its underlying symmetry \cite{Bagchi:2010zz}. This 2d field theory is then used to reproduce gravitational physics in the 3D asymptotically flat spacetime. See \cite{Bagchi:2016bcd, Riegler:2016hah} for a review of important work in this direction.  

\subsection*{Highest weight representations}

The states of the gravitational theory form representations of the underlying symmetry algebra. Again taking a cue out of holography in AdS and also its extensions to higher spins and dS spacetimes, we will consider the highest weight representation of BMS$_3$. We label the states of the dual 2d theory with the centre of the algebra which turns out to be $L_0$ and $M_0$. 
\be{}
L_0 |\D, \xi\> = \D|\D, \xi\>, \quad M_0 |\D, \xi \> = \xi |\D, \xi\>.
\ee
We wish to build on our CFT intuition and thus define BMS primary states as the states $|\D, \xi\> $ which have the lowest value of $\D$ for a given $\xi$. Since acting $L_n$ and $M_{n}$ on  $|\D, \xi\> $ lowers the eigenvalue of $L_0$ by $n$
\be
L_0 L_{n} |\D, \xi\> = (\D-n) L_{n} |\D, \xi\>,\,\,\,L_0 M_{n} |\D, \xi\> = (\D-n) M_{n} |\D, \xi\>,
\ee 
we would impose that 
\be{}
L_n |\D, \xi \> = M_n |\D, \xi \> = 0 \quad \forall n>0.
\ee
It will be particularly convenient (but not essential) to have the BMS analog of the state-operator correspondence and hence we will demand that the states $|\D, \xi\>$  are created by acting the primary field $\phi_{\D,\xi}$ on the vacuum
\be
\phi_{\D,\xi}(0,0)|0\rangle = |\D,\xi\rangle .
\ee
Here the vacuum is the one which is annihilated by the global sub-algebra $\{L_{0, \pm1}, M_{0, \pm1} \}$, which is the Poincare sub-algebra $iso(2,1)$ of BMS$_3$. 

We can increase the eigenvalue of $L_0$ by acting the raising operator $L_{-n}$ and $M_{-n}$ on the BMS primary states. The set of all states obtained from $|\D,
\xi \rangle$ and their linear combination is called the BMS module for $|\D, \xi \rangle$. We will denote this module by $\mathcal{B}(c_L,c_M,\Delta,\xi)$. Then the Hilbert space of the BMS theory is the direct sum of the BMS module of all primaries present in the theory
\be 
\mathcal{H}_{BMS}(c_L,c_M)=\underset{(\Delta,\xi)}{\bigoplus} \mathcal{B}(c_L,c_M,\Delta,\xi).
\ee
States in the module has the general form
\be
L_{-1}^{k_1}L_{-2}^{k_2}....L_{-l}^{k_l} M_{-1}^{q_1}M_{-2}^{q_2}....M_{-r}^{q_r}|\D,\xi \> \equiv L_{\vec{k}}M_{\vec{q}}|\D,\xi \> ,
\ee
where $\vec{k}=(k_1,k_2,....,k_l)$ and $\vec{q}=(q_1,q_2,....,q_r)$ and its $L_0$ eigenvalue is given by
\be
L_0 L_{\vec{k}}M_{\vec{q}}|\D,\xi \> = (N + \Delta) L_{\vec{k}}M_{\vec{q}}|\D,\xi \>, \quad \mbox{where} \ N=\sum_{i}ik_i + \sum_j jq_j.  
\ee 
$N$ is called the level of the state and states in the BMS module are grouped according to their level. For example, level 0 consists of the BMS primary $|\Delta,\xi\rangle$ and level 1 consists of $L_{-1}|\Delta,\xi\rangle$ and $M_{-1}|\Delta,\xi\rangle$. We have given the states upto level 3 in table below.

\bigskip

\begin{table}[h]
\begin{tabular}{|l|l|}
\hline 
Level & States\tabularnewline
\hline 
\hline 
N=0 & $|\Psi_{1}\rangle=|\Delta,\xi\rangle$\tabularnewline
\hline 
N=1 & $|\Psi_{1}\rangle=L_{-1}|\Delta,\xi\rangle,\,\,\,|\Psi_{2}\rangle=M_{-1}|\Delta,\xi\rangle$\tabularnewline
\hline 
N=2 & $\begin{array}{l}
|\Psi_{1}\rangle=L_{-1}^{2}|\Delta,\xi\rangle,\,\,\,|\Psi_{2}\rangle=L_{-2}|\Delta,\xi\rangle,\,\,\,|\Psi_{3}\rangle=L_{-1}M_{-1}|\Delta,\xi\rangle,\,\,\, |\Psi_{4}\rangle = M_{-2}|\Delta,\xi\rangle\\
|\Psi_{5}\rangle=M_{-1}^{2}|\Delta,\xi\rangle
\end{array}$\tabularnewline
\hline 
N=3 & $\begin{array}{l}
|\Psi_{1}\rangle=L_{-1}^3|\Delta,\xi\rangle,\,\,\,|\Psi_{2}\rangle=L_{-1}L_{-2}|\Delta,\xi\rangle,\,\,\,|\Psi_{3}\rangle=L_{-1}^2M_{-1}|\Delta,\xi\rangle,\\
|\Psi_{4}\rangle=L_{-3}|\Delta,\xi\rangle,\,\,\,|\Psi_{5}\rangle=L_{-2}M_{-1}|\Delta,\xi\rangle,\,\,\,|\Psi_{6}\rangle=L_{-1}M_{-2}|\Delta,\xi\rangle,\\
|\Psi_{7}\rangle=M_{-3}|\Delta,\xi\rangle,\,\,\,|\Psi_{8}\rangle=L_{-1}M_{-1}^2|\Delta,\xi\rangle,\,\,\,|\Psi_{9}\rangle=M_{-1}M_{-2}|\Delta,\xi\rangle,|\Psi_{10}\rangle=M_{-1}^3|\Delta,\xi\rangle
\end{array}$\tabularnewline
\hline 
\end{tabular}
\caption{States in the BMS module upto level 3}
\label{BMS_states}
\end{table}

\bigskip

We will denote the number of states at level $N$ by $\widetilde{\dim}_{N}$. This will be equal to the number of partitioning the integer $N$ using two color (corresponding to the $L$'s and $M$'s). Correspondingly, given a state at level $N$, we can separate out the contribution of the $L$'s and $M$'s to the level as
\be
N=n_L+n_M,\,\,\,n_L = k_1+2k_2+....+lk_l,\,\,n_M=q_1+2q_2+...sq_s.
\ee 
Then it can be seen that the number of states at level $N$ is 
\be
\widetilde{\dim}_{N}=\sum_{n_L,n_M,\,n_L+n_M=N} p(n_L)p(n_M) = \sum_{m=0}^{N} p(N-m)p(m),
\label{dimN}
\ee
where $p(n)$ is the number of ways to partition an integer $n$. As $L_{-1}$ and $M_{-1}$ annihilate the vacuum, we will not have any states containing these generator in the BMS module of the vacuum. In this case, the number of states is given by
\bea 
\widetilde{\dim}_{N}({\rm vac})&=&\sum_{n_L,n_M,\,n_L+n_M=N} (p(n_L)-p(n_L-1))(p(n_M)-p(n_M-1)) \cr
&= &\sum_{m=0}^N (p(N-m)-p(N-m-1))(p(m)-p(m-1)).
\label{dimN_vac}
\eea

\medskip

\subsection*{Gram matrix of BMS module}
We can construct a matrix by taking inner products of the states in the BMS module 
\be
K_{ij} = \langle \Psi_i |\Psi_j \rangle. 
\ee
These are the Gram matrices of the BMS module. Here $|\Psi_i\>$ are the states of the BMS module, the first few of which are listed in Table \ref{BMS_states}. The inner product is defined through the Hermiticity properties of the generators
\be
L_n^\dagger = L_{-n}, \quad M_n^\dagger = M_{-n}
\ee
Since states belonging to different levels are orthogonal to each other, this matrix is block diagonal. So, we can study them for each level separately and we use the notation $K^{(N)}$ for the matrix at level $N$. For level 1 we have two states and the Gram matrix $K^{(1)}$ is a $2\times2$ matrix given by
\begin{align}
K^{(1)} = \left[ \begin{array}{cc}
\langle\Delta,\xi|L_{1}L_{-1}|\Delta,\xi\rangle & \langle\Delta,\xi|L_{1}M_{-1}|\Delta,\xi\rangle\\
\langle\Delta,\xi|M_{1}L_{-1}|\Delta,\xi\rangle & \langle\Delta,\xi|M_{1}M_{-1}|\Delta,\xi\rangle
\end{array}   \right] 
&= \left [ \begin{array}{cc}
2\Delta & 2\xi\\
2\xi & 0
\end{array}   \right].
\label{GM_level1}
\end{align} 
For level 2 the Gram matrix $K^{(2)}$ is a $5\times 5$ matrix given by
\bea
K^{(2)} = \left[  \begin{array}{ccccc}
4\Delta(2\Delta+1) & 6\Delta & 4\xi(2\Delta+1) & 6\xi & 8\xi^{2}\\
6\Delta & 4\Delta+6c_L & 6\xi & 4\xi+6c_M & 0\\
4\xi(2\Delta+1) & 6\xi & 4\xi^{2} & 0 & 0\\
6\xi & 4\xi+6c_M & 0 & 0 & 0\\
8\xi^{2} & 0 & 0 & 0 & 0
\end{array} \right].
\label{GM_level2} 
\eea
Note that the matrix $K^{(1)}$ and $K^{(2)}$ has a particular triangular structure i.e., non-zero anti-diagonal elements and all matrix entries on the right hand side of the anti-diagonal line are zero. This is due to the particular way we order the basis states in  Table \ref{BMS_states}. It is helpful to know the structure of the Gram matrix. For example, we can see that due to the triangular structure we have just mentioned, the determinant of $K^{(1)}$ and $K^{(2)}$ are simply given by the product of their anti-diagonal elements. We explore in detail the structure of Gram matrix for general levels in Appendix \ref{appendix_1}. This will in turn be used to find the character of the BMS module in Sec~\ref{innerprod}.

\subsection*{Correlation functions, partition function and modular invariance}

The BMS$_3$ invariant field theories that we propose as dual field theories to asymptotically 3d flat spacetimes, live on its null boundary. These theories are thus defined on a spacetime metric that is degenerate: 
\be
ds^2_{\mathcal{I}^+} = 0 \times du^2 + d\theta^2,
\ee
where $u$ is the null-direction and $\theta$ is the angular co-ordinate at null infinity. The topology of the boundary is ${\rm I\!R}_u \times \mathbb{S}^1$. The vector fields that represent the BMS$_3$ algebra on this boundary are given by: 
\be
L_n = e^{in \theta} \left( \p_\theta - i n u \p_u \right), \quad M_n = e^{in \theta}  \p_u
\ee
We can fix the 2 and 3 point correlation functions of BMS primary operators upto constants by considering the action of only the Poincare sub-algebra $\{L_{0, \pm1}, M_{0, \pm 1} \}$ \cite{Bagchi:2013qva}. The expression for the 2-point function for primary operators with weights $(\Delta, \xi)$ turns out to be \cite{Bagchi:2013qva}
\be
G^{(2)}_{\mbox{\tiny{BMS}}}(\theta_1, u_1, \theta_2, u_2) = C_1 \left( 2 \sin \frac{\theta_{12}}{2} \right)^{-2\Delta} \exp\left({- \xi u_{12} \cot \frac{\theta_{12}}{2}}\right). 
\ee
One can glue together the two ends of the boundary cylinder to define the theory on a torus and here is where the notion of modular invariance comes in. 
We will define the partition function of the BMS field theory by 
\be
{\hat{Z}}_{\textrm{\tiny BMS}} (\sigma, \rho) = {\mbox{Tr}} \,\ e^{2 \pi i \sigma L_0 } e^{2 \pi i \rho M_0} = \sum_{\D, \xi} d(\Delta, \xi) \ e^{2 \pi i \sigma \Delta} e^{2 \pi i \rho \xi},
\ee
where $d(\Delta, \xi)$ denotes the density of states. The BMS version of modular invariance \cite{Bagchi:2013qva} reads
\be{}
\sigma \rightarrow \frac{a \sigma + b}{c \sigma + d} \ , \quad \rho \rightarrow \frac{\rho}{(c \sigma + d)^2} \ .
\ee
This can be derived as a limit from a parent CFT. We will have more to say about this later in the paper. 

The invariance of the partition function under BMS modular transformations, specifically the S-transformation, leads to the BMS-Cardy formula by relating the low-energy spectrum to the high energy spectrum. The BMS-Cardy formula is given by \cite{Bagchi:2012xr}:
\be\label{bmscardy}
S_{\mbox{\tiny{total}}}^{(0)} = \ln d(\Delta, \xi) =  2\pi\bigg( c_L \sqrt{\frac{\xi}{2c_M}} + \Delta \sqrt{\frac{c_M}{2\xi}} \bigg).
\ee   
This is the leading contribution to the entropy in the saddle-point approximation of the integral which gives the density of states. One can also readily compute the next-to-leading piece, which are logarithmically suppressed. This gives the total entropy to be{\footnote{The computations are done in the microcanonical ensemble}}:
\be
S = S_{\mbox{\tiny{total}}}^{(0)} + S_{\mbox{\tiny{total}}}^{(1)} = 2 \pi \left(c_L \sqrt{\frac{\xi}{2c_M}} + \Delta \sqrt{\frac{c_M}{2\xi}}\right) - \frac{3}{2}\log \left(\frac{2\xi}{c_M^{1/3}}\right) + \mbox{constant}
\ee
Further corrections can be found systematically. Actually, the exact integral giving the density of states should be computable analytically following \cite{Loran:2010bd} {\footnote{This observation has been made by Navya Gupta.}}. 

\subsection{3d bulk physics}
We have already stated that a canonical analysis of 3d Einstein gravity with asymptotically flat boundary conditions leads to the BMS$_3$ algebra \refb{bms3} on its null boundary. These asymptotically flat 3d spacetimes are characterised by metrics of the form \cite{Barnich:2010eb, Barnich:2012rz}
\be{}
ds^2 = \Theta(\phi) du^2 - 2 du dr + \left[2 \Xi(\phi) + u \partial_\phi \Theta(\phi)\right] du d\phi + r^2 d\phi^2
\ee
where $u=t-r$ is the retarded time. $\Theta(\phi)$ and $\Xi(\phi)$ are arbitrary functions labelling the solutions called the mass aspect and angular momentum aspect respectively. 

\subsection*{Flat Space Cosmologies}
For the zero mode solutions, the above arbitrary functions just become the mass $(M)$ and the angular momentum $(J)$ upto constants: $\Theta(\phi) = M$ and $\Xi(\phi)=J/2$ and hence the metric takes the form:
\be\label{g1}
ds^2 = M du^2 - 2 du dr + J du d\phi + r^2 d\phi^2
\ee
For non-zero $J$, these solutions are obtained as orbifolds of 3d Minkowski spacetime quotiented by a boost and a translation. The solutions are called shifted-boost orbifolds \cite{Cornalba:2003kd} and also flat space cosmologies (FSC) \cite{Bagchi:2013lma}. These are the analogues of non-extremal BTZ black holes in AdS$_3$. Together with their $J=0$ cousins, the boost orbifold, and the $M=0, J=0$ member, the null orbifold, these form the $M\geq0$ zero mode sector of 3d asymptotically flat gravity. The solution space is depicted in Fig 1(b). For convenience, we also give the zero mode solutions in AdS$_3$ in Fig 1(a){\footnote{We thank an anonymous referee for pointing out errors in a previous version of this figure.}}. In this case, angular deficit is the region bounded by the two lines $\ell M = -|J|$ and the parabola $\ell M=-\frac{l}{8G}-\frac{2G}{l}J^2$, and the range of $J$ given by $|J|\leq \frac{\ell}{4G}$.    

\begin{figure}[h]
\begin{center}
\includegraphics[scale=.4]{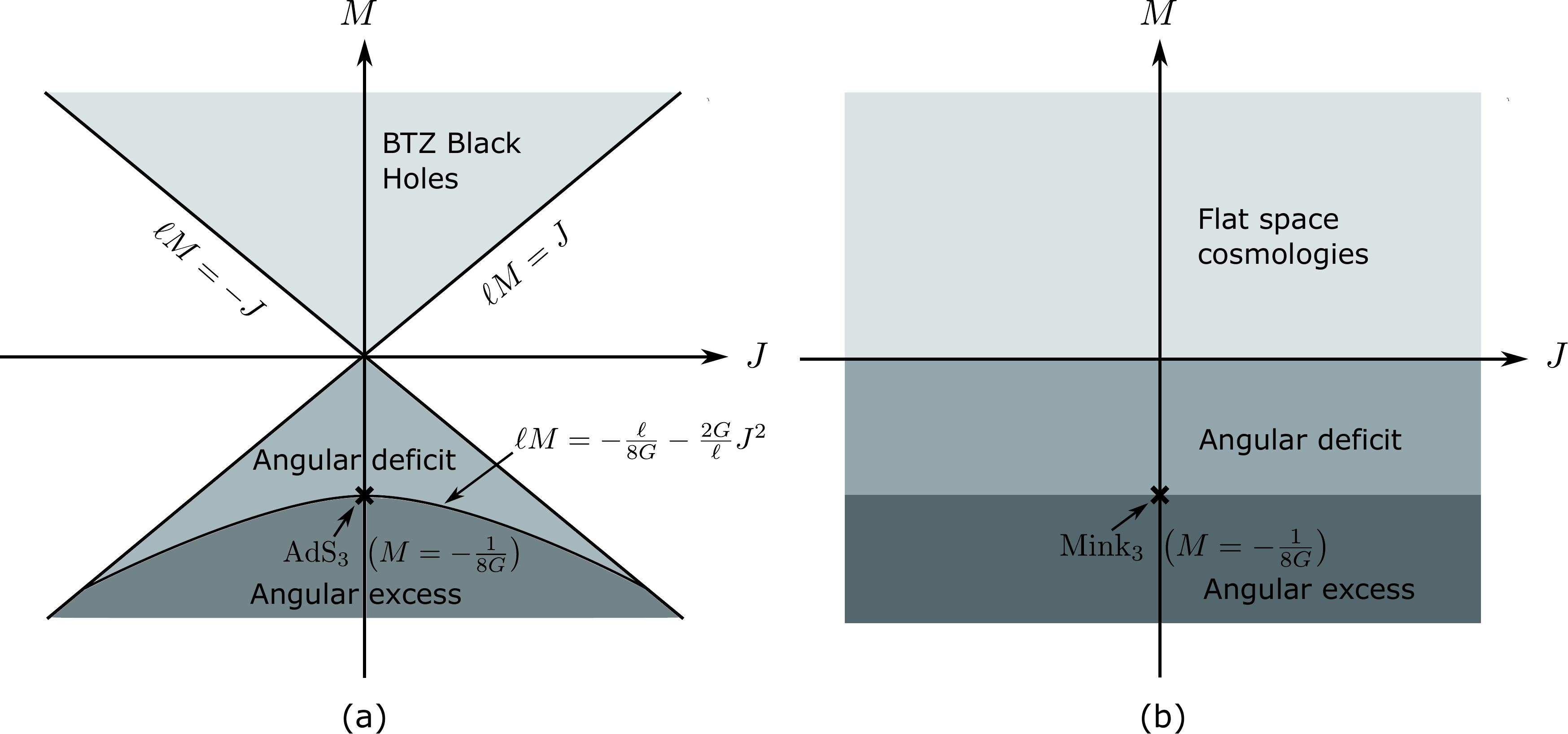}
\end{center}
\caption{Phase space of Einstein gravity in asymptotically (a) AdS$_3$ and (b) Mink$_3$.}
\end{figure}

From the figure above, it seems that FSCs can be understood as a singular limit of the non-extremal BTZ black hole. To understand this, we start off with the global AdS$_3$ metric:
\be{}
ds^2 = - \left(1+ \frac{r^2}{\ell^2}\right) dt^2 +  {\left(1+ \frac{r^2}{\ell^2}\right)}^{-1} {dr^2}+ r^2 d\phi^2
\ee
Here $\ell$ is the AdS radius. To obtain global flat space from AdS, we need to take $\ell \to \infty$. Now consider doing the same on the non-extremal BTZ metric: 
\be{}
ds^2 = - \frac{(r^2 - r_+^2)(r^2 - r_-^2)}{r^2 \ell^2} dt^2 +  \frac{r^2 \ell^2}{(r^2 - r_+^2)(r^2 - r_-^2)}dr^2 + r^2(d\phi - \frac{r_+ r_-}{r^2 \ell} dt)^2.
\ee
Here $r_\pm$ are the outer and inner horizons related to the mass and angular momentum of the BTZ black hole by
\be{}
r_\pm = \sqrt{2G\ell(\ell M +J)} \pm \sqrt{2G\ell(\ell M - J)}
\ee
\begin{figure}[h]
\begin{center}
\includegraphics[scale=.025]{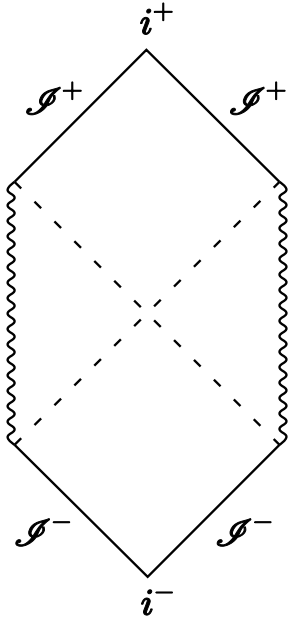}\label{fig}
\end{center}
\caption{Penrose diagram for Flat Space Cosmologies.}
\end{figure}

When we consider the singular flat space limit on the BTZ, we are faced with an apparent conundrum. There are no black holes in 3d flatspace. But following our nose, the solution becomes immediate. When we take $\ell \to \infty$, the outer horizon goes to infinity, while the inner horizon stays put:
\be{}
r_+ \to \sqrt{8GM} \ell = \hat{r}_+ \ell, \quad r_{-} \to \sqrt{\frac{2G}{M}} \ J = r_0
\ee
The resulting metric becomes
\be\label{g2}
ds_{\mbox{\tiny{FSC}}}^2 = \hat{r}_+^2 dt^2 -  \frac{r^2}{\hat{r}_+^2(r^2 - r_0^2)}dr^2 + 2 \hat{r}_+ r_0 d\phi dt + r^2 d\phi^2. 
\ee

The entire spacetime becomes the inside of the outer horizon of the original BTZ black hole. The radial and temporal directions switch roles and hence this is a time dependent cosmological solution. The Penrose diagram of this spacetime is given in Fig.~2. With appropriate rescaling of coordinates, it is easy to see that \refb{g2} reduces to \refb{g1}. 

These FSCs have many interesting properties, the most striking of which is the creation of an FSC by a Hawking-Page like phase transition from empty flat spacetime \cite{Bagchi:2013lma}. Another interesting property, viz. relating FSC on the two separate null boundaries of flat spacetime, $\mathcal{I}^+$ and  $\mathcal{I}^-$,  is explored in \cite{Prohazka:2017equ}.

\subsection*{Entropy and bulk-boundary matching}
FSCs are time-dependent solutions with cosmological horizons. $r=r_0$ is the location of the cosmological horizon and one can associate a Bekenstein-Hawking entropy to it:
\be{}
S_{\mbox{\tiny{FSC}}} = \frac{2 \pi r_0}{4G} = \frac{\pi J}{\sqrt{2GM}}
\ee
For the excited state corresponding to the BTZ blackhole in a 2d CFT, the weights are given by \cite{Kraus:2006wn}
\be
h = \frac{1}{2}  (\ell M + J) + \frac{c}{12}, \quad \h = \frac{1}{2}  (\ell M - J) + \frac{\bar{c}}{12},
\ee
where $c=\bar{c}=\frac{3\ell}{2G}$ is the Brown-Henneaux central charge for the 2d CFT \cite{Brown:1986nw}. Similarly, the BMS weights of the FSC are
\be
\xi = M + \frac{c_M}{2}, \Delta = J + \frac{c_L}{2}
\ee
For 3d Einstein gravity, we have $c_L=0, c_M = 1/4G$. In the limit of large weights, we get 
\be
\xi_{\mbox{\tiny{FSC}}} = M, \quad \Delta_{\mbox{\tiny{FSC}}} = J.
\ee
We plug these values into the BMS-Cardy formula \refb{bmscardy}, and we obtain a perfect matching:
\be
S^{(0)} = 2\pi  \Delta_{\mbox{\tiny{FSC}}} \sqrt{\frac{c_M}{2 \xi_{\mbox{\tiny{FSC}}}}} = \frac{\pi J}{\sqrt{2GM}} = S_{\mbox{\tiny{FSC}}}
\ee
The logarithmic corrections to this also turn out to be of a form expected from the gravitational analysis. For more details, the reader is referred to \cite{Bagchi:2013qva}. We should also point out here that the BMS-Cardy formula has also been derived as a limit from the inner horizon Cardy formula \cite{Castro:2012av} in \cite{Riegler:2014bia,Fareghbal:2014qga}. This has been further used to compute entanglement entropy \cite{Bagchi:2014iea, Basu:2015evh} in BMS-invariant 2d field theories via the so-called Rindler method in \cite{Jiang:2017ecm}. 
\bigskip

\newpage

\section{Highest Weight Characters for BMS}

A torus can be obtained by gluing two ends of a cylinder. We can also twist the cylinder by an angle and then glue the two ends. Then the partition function on a torus twisted by an angle $\theta$ is given by
\be
{\rm Tr}\,e^{-\beta H+i\theta P} ,
\ee
where H is the Hamiltonian which generate transformation along the length of the torus and P generate transformation along the circumference of the torus. If we mapped the cylinder to the plane, we can rewrite $H$ and $P$ in terms of the generator $L_0$ and $M_0$ on the plane
\be
{\rm Tr}\,e^{-\beta H+i\theta P} =  {\rm Tr}\,e^{2\pi i\sigma (L_{0}-c_{L}/2)}e^{2\pi i\rho (M_{0}-c_{M}/2)}\equiv Z_{\mbox{\tiny{BMS}}}(\sigma,\rho).
\ee

\subsection{Formula for character and partition function}
Let us first briefly recall the definition of trace of a linear operator $\hat{O}$ acting on a vector space $V$. If we choose the basis of $V$ to be $|\Psi_i\rangle$, the action of $\hat{O}$ on any basis is given by
\be
\hat{O}|\Psi_i\rangle = \sum_j O_{ij}|\Psi_j\rangle. 
\ee 
Trace of the operator $\hat{O}$ over $V$ is defined as sum of the diagonal elements
\be
{\rm{Tr}}\,\hat{O} \equiv \sum_i O_{ii}.
\label{trace_O} 
\ee
For the BMS partition function, $\hat{O}=e^{2\pi i\sigma (L_{0}-c_L/2)}e^{2\pi i\rho (M_{0}-c_M/2)}$, and $V$ is the Hilbert space $\mathcal{H}_{BMS}(c_L,c_M)$ of the BMS theory. As the operator $\hat{O}$ does not mix states belonging to different BMS modules, we can take trace over each module separately. This give us the character of the module
\bea
\chi_{(c_{L},c_{M},\Delta,\xi)}(\sigma,\rho) & = & {\rm Tr}_{\Delta,\xi}\,e^{2\pi i\sigma (L_{0}-c_{L}/2)}e^{2\pi i\rho (M_{0}-c_{M}/2)},
\eea
where ${\rm Tr}_{\Delta,\xi}$ means trace over the states belonging to the module $ \mathcal{B}(c_L,c_M,\Delta,\xi)$.
The partition function is then given by summing over the characters of the primary fields in the theory
\be
Z_{\mbox{\tiny{BMS}}}(\sigma,\rho) = \sum_{\Delta, \xi} D(\Delta,\xi) \chi_{(c_{L},c_{M},\Delta,\xi)}(\sigma,\rho),
\ee
where $D(\Delta,\xi)$ is multiplicity or density of the primaries with weight $(\Delta,\xi)$.

\medskip

\noindent{\em{Trace at level 1}}

\medskip

\noindent Now, let us try to find an expression for the character $\chi_{(c_{L},c_{M},\Delta,\xi)}(\sigma,\rho)$. As we just stated above, $\hat{O}$ does not mix states with different levels, and hence we will consider the trace for each level separately. Let us first look at level 1 which consists of the states $L_{-1}|\Delta,\xi\rangle$ and $M_{-1}|\Delta,\xi\rangle$. First of all, since all the states in the BMS modules are eigenstates of $L_0$, we have
\bea
&& L_{0}L_{\vec{k}}M_{\vec{q}}|\Delta,\xi\rangle  =(\Delta+N)L_{\vec{k}}M_{\vec{q}}|\Delta,\xi\rangle \\
\Rightarrow &&  e^{2\pi i\sigma L_{0}}L_{\vec{k}}M_{\vec{q}}|\Delta,\xi\rangle  =e^{2\pi i\sigma(N+\Delta)}L_{\vec{k}}M_{\vec{q}}|\Delta,\xi\rangle.
\label{act_L0}
\eea
On the other hand, for $M_0$, we have
\bea
 && e^{2\pi i\rho M_{0}}M_{-1}|\Delta,\xi\rangle = e^{2\pi i\rho \xi}M_{-1}|\Delta,\xi\rangle, \\
&& e^{2\pi i\rho M_{0}}L_{-1}|\Delta,\xi\rangle = (2\pi i\rho)e^{2\pi i\rho\xi}M_{-1}|\Delta,\xi\rangle+e^{2\pi i\rho\xi}L_{-1}|\Delta,\xi\rangle.
\eea
Combining all these we have 
\bea
\hat{O}\left(\begin{array}{c}
L_{-1}|\Delta,\xi\rangle\\
M_{-1}|\Delta,\xi\rangle
\end{array}\right)=e^{-2\pi i(\sigma \frac{c_{L}}{2}+\rho \frac{c_{M}}{2})}e^{2\pi i\sigma(1+\Delta)}e^{2\pi i\rho\xi}\left(
\begin{array}{cc}
1 & 2\pi i\rho \cr
0 & 1
\end{array} 
\right)\left(\begin{array}{c}
L_{-1}|\Delta,\xi\rangle\\
M_{-1}|\Delta,\xi\rangle
\end{array}\right). 
\eea
So, the trace over level 1 is given by
\be
 {\rm Tr}_{\Delta,\xi}\hat{O}_{1} = 2 e^{-2\pi i(\sigma \frac{c_{L}}{2}+\rho \frac{c_{M}}{2})}e^{2\pi i\sigma(1+\Delta)}e^{2\pi i\rho\xi}.
\ee
The factor of two is the number of states that we have in the BMS module at the first level. 
\medskip

\noindent{\em{Trace at general level}}

\medskip

\noindent It is straightforward to find the diagonal elements of the operator $\hat{O}$ for general level. If we take any state $L_{\vec{k}}M_{\vec{q}}|\Delta,\xi\rangle$ at level $N$, the action  of $\hat{O}$ on this state is given by
\bea
\hat{O}L_{\vec{k}}M_{\vec{q}}|\Delta,\xi\rangle &=& e^{-2\pi i(\sigma \frac{c_{L}}{2}+\rho \frac{c_{M}}{2})} e^{2\pi i\sigma L_{0}} e^{2\pi i\rho M_{0}} L_{\vec{k}}M_{\vec{q}}|\Delta,\xi\rangle \nonumber\\
&& \hspace{-2.5cm}= e^{-2\pi i(\sigma \frac{c_{L}}{2}+\rho \frac{c_{M}}{2})}e^{2\pi i\sigma(N+\Delta)}\left([e^{2\pi i\rho M_{0}},L_{\vec{k}}M_{\vec{q}}]|\Delta,\xi\rangle + L_{\vec{k}}M_{\vec{q}}e^{2\pi i\rho M_{0}}|\Delta,\xi\rangle\right) \nonumber\\
&&\hspace{-2.5cm}= e^{-2\pi i(\sigma \frac{c_{L}}{2}+\rho \frac{c_{M}}{2})}e^{2\pi i\sigma(N+\Delta)}\left([e^{2\pi i\rho M_{0}},L_{\vec{k}}M_{\vec{q}}]|\Delta,\xi\rangle + e^{2\pi i\rho \xi}L_{\vec{k}}M_{\vec{q}}|\Delta,\xi\rangle\right)
\label{act_O}
\eea
Since commutator of $M_0$ with $L_n$ changes $L_n$ to $M_n$, i.e. $[M_0,L_n]=-nM_n$,  the states $[e^{2\pi i\rho M_{0}},L_{\vec{k}}M_{\vec{q}}]|\Delta,\xi\rangle$ will not contain $L_{\vec{k}}M_{\vec{q}}|\Delta,\xi\rangle$. Therefore, from \eqref{act_O} we have 
\bea 
\hat{O}L_{\vec{k}}M_{\vec{q}}|\Delta,\xi\rangle &=&  e^{-2\pi i(\sigma \frac{c_{L}}{2}+\rho \frac{c_{M}}{2})}e^{2\pi i\sigma(N+\Delta)}\left(\text{other states}\right) \cr
&& + e^{-2\pi i(\sigma \frac{c_{L}}{2}+\rho \frac{c_{M}}{2})}e^{2\pi i\sigma(N+\Delta)}e^{2\pi i\rho \xi}L_{\vec{k}}M_{\vec{q}}|\Delta,\xi\rangle.
\eea
So, the diagonal elements of $\hat{O}$ are all the same and given by $e^{-2\pi i(\sigma \frac{c_{L}}{2}+\rho \frac{c_{M}}{2})}e^{2\pi i\sigma(N+\Delta)}e^{2\pi i\rho \xi}$. Then the trace at level $N$ is simply given by
\bea
{\rm Tr}_{\Delta,\xi}\,\hat{O}_{N} = \sum_{i=1}^{\widetilde{\dim}_N}  O_{ii} &=&
\sum_{i=1}^{\widetilde{\dim}_N}  e^{-2\pi i(\sigma \frac{c_{L}}{2}+\rho \frac{c_{M}}{2})} e^{2\pi i\sigma(N+\Delta)}e^{2\pi i\rho \xi} \cr &=& \widetilde{\dim}_N \, e^{-2\pi i(\sigma \frac{c_L}{2}+\rho \frac{c_M}{2})} e^{2\pi i\sigma(N+\Delta)}e^{2\pi i\rho \xi},
\label{trace_level_N}
\eea
where $\widetilde{\dim}_N$ is the number of linearly independent descendant states at level $N$.
Hence
\bea
 \chi_{(c_{L},c_{M},\Delta,\xi)}(\sigma,\rho) &=&  \sum_N {\rm Tr}_{\Delta,\xi}\,\hat{O}_N =e^{-2\pi i(\sigma \frac{c_{L}}{2}+\rho \frac{c_{M}}{2})}e^{2\pi i(\sigma\Delta+\xi\rho)}\sum_{N}\widetilde{\dim}_{N}e^{2\pi i\sigma N}.
 \label{BMS_character}
\eea
Substituting the expression for $\widetilde{\dim}_{N}$ given by \eqref{dimN} in the above formula, we have
\bea
\chi_{(c_{L},c_{M},\Delta,\xi)}(\sigma,\rho) &=& e^{-2\pi i(\sigma \frac{c_{L}}{2}+\rho \frac{c_{M}}{2})}e^{2\pi i(\sigma\Delta+\xi\rho)}\sum_{N}\sum_{n_L+n_M=N} p(n_L) p(n_M)e^{2\pi i\sigma (n_L+n_M)}\cr
\Rightarrow \chi_{(c_{L},c_{M},\Delta,\xi)} &=& e^{-2\pi i(\sigma \frac{c_{L}}{2}+\rho \frac{c_{M}}{2})}e^{2\pi i(\sigma\Delta+\xi\rho)} \sum_{n_L} p(n_L) e^{2\pi i\sigma n_L}  \sum_{n_M} p(n_M) e^{2\pi i\sigma n_M}.  
\label{BMS_character2} 
\eea
Using the generating function of partition numbers and the definition of the Dedekind eta function
\be
\prod_{n=1}^{\infty}\frac{1}{1-x^n}=\sum_{n=0}^{\infty} p(n) x^n, \quad \eta(\sigma)=e^{\frac{2\pi i \sigma}{24}}\prod_{n=1}^{\infty}(1-e^{2\pi i \sigma n}),
\ee
we can rewrite \eqref{BMS_character2} as 
\be
\chi_{(c_{L},c_{M},\Delta,\xi)}(\sigma,\rho) = \frac{e^{\frac{2\pi i \sigma}{12}} e^{-2\pi i(\sigma \frac{c_L}{2}+\rho \frac{c_M}{2})}e^{2\pi i(\sigma\Delta+\xi\rho)}}{\eta(\sigma)^2}.
\label{BMS_character3}
\ee
The above formula is for non-vacuum states. 
For the vacuum $(\Delta=0,\xi=0)$, we have to use \eqref{dimN_vac} for number of states  giving us
\bea
\chi_{(c_{L},c_{M},0,0)}(\sigma,\rho) &=&  \frac{e^{\frac{2\pi i \sigma}{12}} e^{-2\pi i(\sigma \frac{c_{L}}{2}+\rho \frac{c_{M}}{2})}}{\eta(\sigma)^2}(1-e^{2\pi i \sigma})^2.
\eea
Finally, the partition function is given by
\bea
 Z_{\mbox{\tiny{BMS}}}(\sigma,\rho) &&= \sum_{\Delta,\xi}  D(\Delta,\xi) \, \chi_{(c_{L},c_{M},\Delta,\xi)}(\sigma,\rho) \\
&& =\frac{e^{\frac{2\pi i \sigma}{12}} e^{-2\pi i(\sigma \frac{c_{L}}{2}+\rho \frac{c_{M}}{2})}}{\eta(\sigma)^2} \left(\tilde{D}(0,0)(1-e^{2\pi i \sigma})^2+\sum_{\Delta, \xi \neq 0}  D(\Delta,\xi) e^{2\pi i(\sigma\Delta+\rho\xi)}\right)\nonumber
\label{partition_function}
\eea
where we use the notation $\tilde{D}(0,0)$ for the density of vacuum state. If the vacuum is non-degenerate we can simply take this to be 1. If we introduce $D(0,0)=\tilde{D}(0,0)(1-e^{2\pi i \sigma})^2$, the above formula can be rewritten as
\bea
Z_{\mbox{\tiny{BMS}}}(\sigma,\rho) &=& \frac{e^{\frac{2\pi i \sigma}{12}} e^{-2\pi i(\sigma \frac{c_{L}}{2}+\rho \frac{c_{M}}{2})}}{\eta(\sigma)^2} \sum_{\Delta, \xi}  D(\Delta,\xi) e^{2\pi i(\sigma\Delta+\rho\xi)}.
\label{full_partition}
\eea

\bigskip

\subsection{Character in terms of inner product}\label{innerprod}
If a vector space is equipped with an inner product, we can write the trace of an operator using this inner product. Let $K_{ij}$ be the Gram matrix formed from the inner product of the basis states $|\Psi_i\rangle$'s
\be
K_{ij}=\langle \Psi_i|\Psi_j\rangle, 
\ee 
then trace of an operator $\hat{O}$ is given by
\be
{\rm Tr}\,\hat{O} = \sum_{i,j} K^{ij}\langle \Psi_i|\hat{O}|\Psi_j\rangle,  
\label{tr_2}
\ee
where $K^{ij}$ is the matrix inverse of $K_{ij}$. It is easy to check that this is independent of the basis we use. 

Now, let us use \eqref{tr_2} to express the character of the BMS module $\mathcal{B}(c_L,c_M,\Delta,\xi)$.  For the character, the operator $\hat{O}$ is once again $e^{2\pi i\sigma (L_{0}-c_L/2)}e^{2\pi i\rho (M_{0}-c_M/2)}$. Since states in different levels are orthogonal to each other we may write 
\be
 \chi_{(c_{L},c_{M},\Delta,\xi)}(\sigma,\rho) = \sum_{i,j,N} K^{ij}_{(N)}\langle \Psi_i^{(N)}|\hat{O}|\Psi_j^{(N)}\rangle \equiv  \sum_{i,j,N} K^{ij}_{(N)}\tilde{O}_{ij}^{(N)},
\ee 
where $K_{(N)}$ is the inverse of the level $N$ Gram matrix $K^{(N)}$ and the $|\Psi_i^{(N)}\rangle$'s are basis states of the BMS module at level $N$. So, we can calculate the trace for each level separately and then add everything at the end. We will calculate this for level 1 and level 2 below and give a general proof for arbitrary level in Appendix \ref{appendix_2}.

For level 1 we have two basis states (see Table \eqref{BMS_states}) with the Gram matrix $K^{(1)}$ given in \eqref{GM_level1}. The inverse matrix $K_{(1)}$ is given by
\begin{align}
K_{(1)} = \left[
\begin{array}{cc}
 0 & \frac{1}{2 \xi } \\
 \frac{1}{2 \xi } & -\frac{\Delta }{2 \xi ^2} \\
\end{array}
\right].
\end{align}
For $\tilde{O}^{(1)}$ we have
\be
\tilde{O}^{(1)} =  e^{-2\pi i(\sigma \frac{c_L}{2}+\rho \frac{c_M}{2})} \left[ \begin{array}{cc}
\langle L_{1}e^{2\pi i\sigma L_{0}}e^{2\pi i\rho M_{0}}L_{-1}\rangle & \langle L_{1}e^{2\pi i\sigma L_{0}}e^{2\pi i\rho M_{0}}M_{-1} \rangle\\
\langle M_{1}e^{2\pi i\sigma L_{0}}e^{2\pi i\rho M_{0}}L_{-1} \rangle & \langle M_{1}e^{2\pi i\sigma L_{0}}e^{2\pi i\rho M_{0}}M_{-1}\rangle
\end{array} \right]\nonumber\\
\ee
where $\langle \ldots \rangle = \langle\Delta,\xi| \ldots |\Delta,\xi\rangle$. Hence we have  
\be
\tilde{O}^{(1)} =  e^{-2\pi i(\sigma \frac{c_L}{2}+\rho \frac{c_M}{2})} e^{2\pi i\sigma (\Delta+1)}e^{2\pi i\rho \xi}\left[  \begin{array}{cc}
(2\pi i\rho)2\xi+2\Delta &  \ 2\xi\\
2\xi & \ 0
\end{array}  \right].
\ee
Then the trace over level 1 is
\be
\sum_{i,j}K^{ij}_{(1)}\tilde{O}_{ij}^{(1)} = 2  e^{-2\pi i(\sigma \frac{c_L}{2}+\rho \frac{c_M}{2})} e^{2\pi i\sigma (\Delta+1)}e^{2\pi i\rho \xi}.
\ee
Here the pre-factor 2 is the number of states at level 1.  

Now let us calculate the trace for level 2. Here we have five descendant states (see Table \eqref{BMS_states}) and from \eqref{GM_level2} we can see that $K^{(2)}$ has a triangular structure with non-zero anti-diagonal elements and all the matrix elements on the right hand side of the anti-diagonal line being zero. Then we can see that $K_{(2)}$ will have opposite structure
\begin{align}
K_{(2)} = \left[
\begin{array}{ccccc}
 0 & 0 & 0 & 0 & \frac{1}{8 \xi ^2} \\
 0 & 0 & 0 & \frac{1}{4 \xi +6 c_M} & K^{25} \\
 0 & 0 & \frac{1}{4 \xi ^2} & K^{34} & K^{35} \\
 0 & \frac{1}{4 \xi +6 c_M} & K^{43} & K^{44} & K^{45} \\
 \frac{1}{8 \xi ^2} & K^{52} & K^{53} & K^{54} & K^{55} \\
\end{array}
\right],
\end{align}
with all matrix elements on the left hand side of the anti-diagonal line being zero and the anti-diagonal elements being the inverse of that of $K^{(2)}$
\be
K^{ii}_{(2)} =  \frac{1}{K_{ii}^{(2)}}.
\label{rel_1}
\ee
As for the matrix $\tilde{O}^{(2)}$ we have
\bea
\tilde{O}^{(2)}=   e^{-2\pi i(\sigma \frac{c_L}{2}+\rho \frac{c_M}{2})} e^{2\pi i\sigma (\Delta+2)}e^{2\pi i\rho \xi} \left [ \begin{array}{ccccc}
\tilde{O}_{11} & \tilde{O}_{12} & \tilde{O}_{13} & \tilde{O}_{14} & 8\xi^{2}\\
\tilde{O}_{21} & \tilde{O}_{22} & \tilde{O}_{23} & 4\xi+6c_M & 0\\
\tilde{O}_{31} & \tilde{O}_{22} & 4\xi^{2} & 0 & 0\\
\tilde{O}_{41} & 4\xi+6c_M & 0 & 0 & 0\\
8\xi^{2} & 0 & 0 & 0 & 0
\end{array} \right].
\eea
Note that this matrix is triangular in the same way as the Gram matrix $K^{(2)}$ and its anti-diagonal elements are proportional to that of $K^{(2)}$ by a common factor 
\be
\tilde{O}_{ii}^{(2)} = e^{-2\pi i(\sigma \frac{c_L}{2}+\rho \frac{c_M}{2})} e^{2\pi i\sigma (\Delta+2)}e^{2\pi i\rho \xi} K_{ii}^{(2)}.   
\label{rel_2}
\ee 
Since the matrix $\tilde{O}^{(2)}$ and $K_{(2)}$ are triangular in the opposite way, only the anti-diagonal elements of both matrix will contribute to the trace $\sum_{i,j}K^{ij}_{(2)}\tilde{O}_{ij}^{(2)}$. Using this information along with \eqref{rel_1} and \eqref{rel_2} we have
\bea
\sum_{i,j}K^{ij}_{(2)}\tilde{O}_{ij}^{(2)} &=& \sum_{i=1}^{5} K^{ii}_{(2)}\tilde{O}_{ii}^{(2)} = e^{-2\pi i(\sigma \frac{c_L}{2}+\rho \frac{c_M}{2})} e^{2\pi i\sigma (\Delta+2)}e^{2\pi i\rho \xi} \sum_{d=1}^{5} 1\cr
&=& 5 e^{-2\pi i(\sigma \frac{c_L}{2}+\rho \frac{c_M}{2})} e^{2\pi i\sigma (\Delta+2)}e^{2\pi i\rho \xi} ,
\eea
where the sum, $\sum_d 1$, just gave us number of states at level 2 which is 5. So, we can already guess that for arbitrary level $N$ 
\be
\sum_{i,j}K^{ij}_{(N)}\tilde{O}_{ij}^{(N)} =  \widetilde{\rm dim}_N \,e^{-2\pi i(\sigma \frac{c_L}{2}+\rho \frac{c_M}{2})} e^{2\pi i\sigma (\Delta+N)}e^{2\pi i\rho \xi}, 
\label{trace_level_N_2}
\ee
in agreement with \eqref{trace_level_N}. The upper/lower triangular structure of the above matrices don't survive to higher order. But one can systematically organise them to arrive at the above expression for the trace at a general level. We will prove this in Appendix \ref{appendix_2}. Using the above formula, the character is the same as  \eqref{BMS_character}
\be
 \chi_{(c_{L},c_{M},\Delta,\xi)}(\sigma,\rho) = \sum_{i,j,N}K^{ij}_{(N)}\tilde{O}_{ij}^{(N)} =  e^{-2\pi i(\sigma \frac{c_L}{2}+\rho \frac{c_M}{2})} e^{2\pi i(\sigma \Delta + \rho \xi)} \sum_N \widetilde{\rm dim}_N \, e^{2\pi i\sigma N}.
\ee

\bigskip

\section{Limiting analysis from character of 2d CFT}
In this section we will re-derive our formula for BMS character \eqref{BMS_character} by taking limits on the 2d CFT character. 

\subsection{The Virasoro character} 
The holomorphic part of the character for the Virasoro module generated by the primary field with conformal weight $(h,\bar{h})$ is given by
\begin{align}
\chi_{(c,h)}(\tau)& ={\rm Tr}_h\,q^{\mathcal{L}_{0}-c/24}.
\end{align}
where $q=e^{2\pi i\tau}$. 
This is easy to calculate as all the states of the Virasoro module 
\be
 \mathcal{L}_{\vec{k}}|h\rangle \equiv (\mathcal{L}_{-1})^{k_{1}}(\mathcal{L}_{-2})^{k_{2}}...(\mathcal{L}_{-r})^{k_{r}} |h\rangle
\ee
are eigenstates of $\mathcal{L}_0$ 
\be
\mathcal{L}_0  \mathcal{L}_{\vec{k}}|h\rangle = (n+h) \mathcal{L}_{\vec{k}}|h\rangle,\,\,\,n=\sum_l lk_l.
\ee
The number of states with eigenvalue $(h+n)$ (except for $h=0$) is given by the partition number $p(n)$. So we have
\be 
\chi_{(c,h)}(\tau)=\sum_n p(n)q^{(n+h)-c/24}.
\ee
 We have the same structure for the anti-hlomorphic part. The character for $(h,\bar{h})$ is then given by
\begin{eqnarray}
\chi_{(c,h)}(\tau)\chi_{(\bar{c},\bar{h})}(\bar{\tau}) 
 & = & q^{-c/24}\bar{q}^{-\bar{c}/24}\sum_{n}p(n)q^{n+h}\sum_{\bar{n}}p(\bar{n})\bar{q}^{\bar{n}+\bar{h}}.
\label{eq:cft_character}
\end{eqnarray}
For the vacuum $(h=0,\bar{h}=0)$, we will not have the states containing $\mathcal{L}_{-1}(\bar{\mathcal{L}}_{-1})$ as these generators annihilate the vacuum state. So, the character for this case will be given by
\begin{eqnarray}
\chi_{(c,0)}(\tau)\chi_{(\bar{c},0)}(\bar{\tau}) = q^{-c/24}\bar{q}^{-\bar{c}/24}\sum_{n}(p(n)-p(n-1))q^{n}\sum_{\bar{n}}(p(\bar{n})-p(\bar{n}-1))\bar{q}^{\bar{n}}.
\end{eqnarray}

\subsection{The two limits and the limiting characters} 
We expect to get the character of the BMS module by taking limit on the Virasoro character. We will do this analysis for non-vacuum primaries for both the non-relativistic (NR) and the ultra-relativistic (UR) limits. We first remind the reader of the two different contraction that gets one from the two copies of the Virasoro algebra to the BMS$_3$. The first one is the non-relativistic contraction:
\be{}
\mbox{NR limit:} \hspace{1cm} L_n = \L_n + \bcL_n, \quad M_n = -\e (\L_n - \bcL_n) \label{NR}
\ee
The name is derived from the fact that if one looks at the generators of the Virasoro algebra on the cylinder and takes a spacetime contraction where the speed of light is taken to infinity, these are the linear combination of generators required to give finite answers. One can also take the Carrollian ($c\to 0$) or the ultra-relativistic limit instead of the NR limit. In this case, the linear combination of generators are:
\be{}
\mbox{UR limit:} \hspace{1cm} L_n = \L_n - \bcL_{-n}, \quad M_n = \e (\L_n + \bcL_{-n})  \label{UR}
\ee
The mappings of the central terms, the weights and the modular parameters from the relativistic to the NR/UR theory are given in the equations below: 
\begin{subequations}
\begin{align}
&{\mbox{NR limit:}} \quad  (c, \bar{c}) = 6 \left(c_{L}\mp\frac{c_{M}}{\epsilon}\right); \ (h,\bar{h}) = \frac{1}{2}\left(\Delta \mp \frac{\xi}{\epsilon}\right); \
(\tau, \bar{\tau}) = \pm \sigma-\epsilon\rho; \\
&{\mbox{UR limit:}} \quad  (c, \bar{c}) = 6 \left(\pm c_{L} + \frac{c_{M}}{\epsilon}\right); \ (h,\bar{h}) = \frac{1}{2}\left(\pm \Delta + \frac{\xi}{\epsilon}\right); \ (\tau, \bar{\tau}) = \sigma \pm \epsilon\rho.
\end{align}
\end{subequations}
In the NR limit, the CFT character \eqref{eq:cft_character} reduces to 
\begin{eqnarray}
 \chi^{NR}(\sigma,\rho)& = & \lim_{\e\to0} \ q^{-c/24}\bar{q}^{-\bar{c}/24}\sum_{n}p(n)q^{n+h}\sum_{\bar{n}}p(\bar{n})\bar{q}^{\bar{n}+\bar{h}}\cr
 & = &  \lim_{\e\to0} \ e^{-2\pi i(\sigma-\epsilon\rho)\frac{6}{24}(c_L-\frac{c_M}{\epsilon})}e^{-2\pi i(\sigma+\epsilon\rho)\frac{6}{24}(c_L+\frac{c_M}{\epsilon})}\cr
 &&\times \sum_{n}p(n)e^{2\pi i(\sigma-\epsilon\rho)(n+\frac{1}{2}(\Delta-\frac{\xi}{\epsilon}))}\sum_{\bar{n}}p(\bar{n})e^{2\pi i(\sigma+\epsilon\rho)(\bar{n}+\frac{1}{2}(\Delta+\frac{\xi}{\epsilon}))}\cr
\implies \chi^{NR}(\sigma,\rho) &= &  e^{-2\pi i(\sigma \frac{c_L}{2}+\rho \frac{c_M}{2})}e^{2\pi i(\sigma\Delta+\xi\rho)} \frac{1}{|\phi(\sigma)|^2}
\label{NRchar}.
\end{eqnarray}
where $\phi(\sigma)=\prod_{n=1}^{\infty}(1-e^{2\pi i \sigma})$.
Similarly, for the UR limit, we have
\begin{eqnarray}
 \chi^{UR}(\sigma,\rho)& = & \lim_{\e\to0} \ q^{-c/24}\bar{q}^{-\bar{c}/24}\sum_{n}p(n)q^{n+h}\sum_{\bar{n}}p(\bar{n})\bar{q}^{\bar{n}+\bar{h}}\cr
 & = & \lim_{\e\to0} \ e^{-2\pi i(\sigma+\epsilon\rho)\frac{6}{24}(c_L+\frac{c_M}{\epsilon})}e^{-2\pi i(\sigma-\epsilon\rho)\frac{6}{24}(-c_L+\frac{c_M}{\epsilon})}\cr
 &&\times \sum_{n}p(n)e^{2\pi i(\sigma+\epsilon\rho)(n+\frac{1}{2}(\Delta+\frac{\xi}{\epsilon}))}\sum_{\bar{n}}p(\bar{n})e^{2\pi i(\sigma-\epsilon\rho)(-\bar{n}+\frac{1}{2}(\Delta-\frac{\xi}{\epsilon}))}\cr
\implies \chi^{UR}(\sigma,\rho) & = & e^{-2\pi i(\sigma \frac{c_L}{2}+\rho \frac{c_M}{2})}e^{2\pi i(\sigma\Delta+\xi\rho)}\frac{1}{|\phi(\sigma)|^2}.
\end{eqnarray}
The characters in the limit, viz. $\chi^{NR}(\sigma,\rho), \chi^{UR}(\sigma,\rho)$ are the same and are identical to what we have obtained using intrinsic methods in \eqref{BMS_character3} {\footnote{We should point out that \eqref{BMS_character3} does not contain the mod-squared as the limiting characters. But the equality still holds. This is because in the intrinsic character, which is also follows from the NR limit, the parameter $\s$ is purely imaginary, thus making $\eta(\s)$ purely real. The UR limit is more subtle and the reader is referred to \cite{Oblak:2015sea} for further details on this.}}.

\subsection{Equal limiting characters: the problem} 
At the outset, the fact that the characters in the two limits are the same and reproduce the answer in the intrnisic analysis is not surprising given that both limits brought us to the same algebra from the two copies of the Virasoro. But there is something deeply profound about the above statements. We shall try to address why the matching of answers in the two limits is extremely surprising and then go on to give some partial answer to the puzzle. 

Characters are properties of the representations of a particular algebra and not the algebra per se. For finite groups, the characters of a certain representation are the trace of representative matrix. If two representations have equal characters, we know that these representations are linked by similarity transformations. The Virasoro characters \refb{eq:cft_character} are constructed for the highest weight representations that are characterised by weights $(h, \bar{h})$. As a reminder, highest weight representations of the Virasoro are built on primary states that defined as
\be\label{vhw}
\L_0 | h,  \bar{h}\rangle = h | h,  \bar{h}\rangle, \ \bcL_0 | h,  \bar{h}\rangle = \bar{h} | h,  \bar{h}\rangle; \quad \L_n | h,  \bar{h}\rangle = 0, \bcL_n | h,  \bar{h}\rangle = 0 \ \ \forall n >0.
\ee
The Virasoro modules are built by acting raising operators on these states. 
What about the BMS characters? In the first part of the paper, we constructed these in terms of highest weight representations as well. 

So far, there seems to be no trouble. But let's look at the two limits. The NR limit \refb{NR} maps Virasoro highest weight states to BMS highest weight states. 
\be{}
\refb{vhw} \rightarrow L_0 |\D, \xi \> = \D |\D, \xi\>,  M_0 |\D, \xi \> = \xi |\D, \xi\>; \ L_n |\D, \xi\>= 0 = M_n |\D, \xi\> \  \forall n >0. 
\ee
Hence the fact that we reproduced the BMS highest weight characters as a NR limit of the Virasoro characters in \refb{NRchar} is very natural, and is a robust check of our previous analysis. 

On the other hand, due to the mixing of positive and negative modes in the linear combination, the UR limit \refb{UR} definitely does not take the Virasoro highest weights to BMS highest weights. This in fact maps to something call the BMS induced representations \cite{Barnich:2014kra, Barnich:2015uva,Campoleoni:2016vsh, Oblak:2016eij}, which are defined by:
\be
\refb{vhw} \rightarrow L_0 |\D, \xi \> = \D |\D, \xi\>, M_0 |\D, \xi\> = \xi |\D, \xi\>; \quad M_n  |\D, \xi\> = 0 \ \forall n \neq 0. 
\ee
These representations are clearly different from the highest weight representations that we have talked about so far. In fact, the characters for these representations have been computed by group theoretic methods in \cite{Oblak:2015sea} (see also \cite{Garbarz:2015lua}).  The UR limit thus gives us the character of the induced representation. 

The disturbing aspect of the statements made above is the fact that the characters in the NR and UR limits are one and the same. This means that for the BMS$_3$ algebra, the highest weight representation and the induced representation have identical characters. What makes things even more disturbing is the question of unitarity of the representations. The major advantage of the induced representations is the fact that these are manifestly unitary \cite{Barnich:2014kra, Barnich:2015uva,Campoleoni:2016vsh, Oblak:2016eij}. The highest weight representations, on the other hand, although extremely useful in various holographic applications, are explicitly non-unitary. A quick look at the equivalent of the Kac determinant at level one is enough to convince one of this.  The Gram matrix at the first level is given by \refb{GM_level1}. The determinant is 
\be
\det K^{(1)} = - 4 \xi^2. 
\ee
This is negative for all real values of $\xi$ and hence the representation is non-unitary for all $\xi>0$. The only chance of unitarity is $\xi=0$. It can be shown that this also needs to be combined with $c_M=0$ \cite{Grumiller:2014lna}. But one can also show that when one considers the sub-sector with $(\xi=0, c_M=0)$, there is a truncation of the algebra from BMS$_3$ to a single copy of the Virasoro algebra \cite{Bagchi:2009pe}. One of the principle reasons we are interested in the BMS$_3$ because of the potential connection to holographic physics in flatspace. For that $c_M\neq0$. Hence the $(\xi=0, c_M=0)$ is not a very interesting sub-sector{\footnote{If, however, we venture beyond Einstein gravity, there are interesting holographic duals to this theory. See \cite{Bagchi:2012yk} for the bosonic theory and \cite{Bagchi:2018ryy} for a recent fermionic generalisation.}}. 

So, we have seen that in a generic BMS-invariant theory, the highest weight representation is non-unitary whereas the induced representations are constructed to be unitary. Their characters are however identical. This is a source of great intrigue. 

\subsection{Hints of a solution}
One might assume that the very unexpected mapping between the two apparently inequivalent representations is inherently a property of the BMS$_3$ algebra, and this happens because of the vagaries of the limiting procedure from the 2d CFT. It has been argued that the reason that the NR and the UR limits give the same algebra starting out from two copies of the Virasoro is because there are only two directions in the 2d field theory and the process of contraction is blind to this. 

But there is something more fundamental about this weird NR $\leftrightarrow$ UR mapping. This actually is not a property of the limit, but a strange intrinsic property of 2d CFT itself. Notice that the following operation:
\be{}
\L_n \to - \L_{-n}, \quad c\to -c
\ee
is an automorphism of the Virasoro algebra{\footnote{The automorphism in the Witt algebra (i.e. without the central term) has been noticed before in \cite{Oblak:2016eij}.}}. Hence for two copies of the Virasoro algebra, if we perform the above operation on the anti-holomorphic sector, we get the following automorphism
\be
\L_n + \bcL_n \to \L_n - \bcL_{-n}, \quad \L_n - \bcL_n \to \L_n + \bcL_{-n}, \quad c\pm\bar{c} \to c\mp\bar{c}
\ee
Without the factors of $\e$, this is precisely the NR $\leftrightarrow$ UR swapping. This exchanges 
\be
\L_0 + \bcL_0 \to \L_0 - \bcL_0
\ee
Hence the usual highest weight representation theory of the 2d CFT gets mapped to one where the $\bar{h}$ eigenvalue is not bounded from below, but bounded from above. The characters of these very different representations are the same because of this automorphism. We believe that the route to understanding the conflicting issues in the BMS representation theory is to carefully study this particular automorphism in the parent 2d CFT. 

\bigskip 
\section{BMS-Cardy formula for primary states}
We have obtained the characters for the BMS$_3$ algebra in the previous sections by a variety of methods. As we emphasised in the introduction and later in Sec 2, the BMS$_3$ algebra is central to understanding holography in 3d asymptotically flat spacetimes. Using our expressions for the characters found in the earlier sections, in this section we will obtain a Cardy like formula for density of primaries $D(\Delta,\xi)$ with large $\Delta$ and $\xi$. 

We have in Sec~3 used the characters to write down the partition function of the BMS invariant field theory. In this section, we will use this and the modular invariance of the partition function $Z_{\mbox{\tiny{BMS}}}(\sigma,\rho)$ to arrive at an expression for the entropy of primary states. From \eqref{full_partition}, we have
\be
Z_{\mbox{\tiny{BMS}}}(\sigma,\rho) = \frac{e^{\frac{2\pi i \sigma}{12}} e^{-2\pi i(\sigma \frac{c_L}{2}+\rho \frac{c_M}{2})}}{\eta(\sigma)^2} \tilde{Z}(\sigma,\rho)
\label{partition_tilde}
\ee
where $\tilde{Z}(\sigma,\rho) = \sum_{\Delta,\xi} D(\Delta,\xi) e^{2\pi i(\sigma\Delta+\rho\xi)}$. We can invert the above formula to give us the density of the primary states
\be
D(\Delta,\xi) = \int d\sigma d\rho  e^{-2\pi i(\sigma\Delta+\rho\xi)} \tilde{Z}(\sigma,\rho).
\label{density_ps}
\ee
Using the modular properties of the partition function and Dedekind eta function
\be
Z_{\mbox{\tiny{BMS}}}(\sigma,\rho)=Z_{\mbox{\tiny{BMS}}}\left(-\frac{1}{\sigma},\frac{\rho}{\sigma^2}\right), \quad \sqrt{-i\sigma}\,\eta(\sigma) = \eta\left(-\frac{1}{\s}\right),
\ee
we can deduce that
\bea
\tilde{Z}\left(-\frac{1}{\sigma},\frac{\rho}{\sigma^2}\right) &=& Z_{\mbox{\tiny{BMS}}}\left(-\frac{1}{\sigma},\frac{\rho}{\sigma^2}\right)\eta\left(-\frac{1}{\sigma}\right)^2e^{2\pi i(-\frac{1}{\sigma}\frac{6c_L-1}{12}+\frac{\rho}{\sigma^2} \frac{c_M}{2})}\nonumber \\
\implies \tilde{Z}(\sigma,\rho) &=&  \tilde{Z}\left(-\frac{1}{\sigma},\frac{\rho}{\sigma^2}\right)\left(\frac{i}{\sigma}\right)e^{2\pi i(\frac{1}{\sigma}\frac{6c_L-1}{12}-\frac{\rho}{\sigma^2} \frac{c_M}{2}+\sigma \frac{6c_L-1}{12}+\rho \frac{c_M}{2})}.
\eea
Substituting this in \eqref{density_ps}, we have
\bea
D(\Delta,\xi) &=& \int d\sigma d\rho \, \tilde{Z}\left(-\frac{1}{\sigma},\frac{\rho}{\sigma^2}\right)\left(\frac{i}{\sigma}\right)e^{f(\sigma,\rho)},
\label{density_int}
\eea
where
\be
f(\sigma,\rho)= 2\pi i\left(\frac{1}{\sigma}\frac{6c_{L}-1}{12}-\frac{\rho}{\sigma^2} \frac{c_{M}}{2}+\sigma \frac{6c_{L}-1}{12}+\rho \frac{c_{M}}{2}-\Delta \sigma -\rho \xi\right).
\ee
For large $(\Delta, \xi)$, we employ the saddle-point approximation to calculate the integral \eqref{density_int}: 
\be
 D(\Delta,\xi) \approx \tilde{Z}\left(-\frac{1}{\sigma_c},\frac{\rho_c}{\sigma^2_c}\right)\left(\frac{i}{\sigma_c}\right)e^{f(\sigma_c,\rho_c)},
 \label{density_saddle}
\ee 
where $(\sigma_c,\rho_c)$ is the critical point of the function $f(\sigma,\rho)$: $\partial_{\sigma}f(\sigma_c,\rho_c)=0, \, \partial_{\rho}f(\sigma_c,\rho_c)=0.$
These are given by
\be
-\frac{6 c_L-1}{12 \sigma^2_c}+\frac{1}{12} \left(6 c_L-1\right)+\frac{\rho_c  c_M}{\sigma^3_c}-\Delta = 0, \quad -\frac{c_M}{2 \sigma^2_c}+\frac{c_M}{2}-\xi = 0 .
\ee
This has two solutions which we denote by $(\sigma_0,\rho_0)$ and $(\tilde{\sigma}_0,\tilde{\rho}_0)$. For large $\Delta$ and $\xi$, they are given by
\be
\sigma_{0} \approx i\sqrt{\frac{c_M}{2\xi}},\,\,\,\,\rho_0\approx i\frac{ \left(6 \xi  c_L - 6 \Delta  c_M - \xi \right)}{6 (2 \xi )^{3/2} \sqrt{c_M}},
\ee
\be
\tilde{\sigma}_0 \approx -i\sqrt{\frac{c_M}{2\xi}},\,\,\,\,\tilde{\rho}_0\approx -i\frac{ \left(6 \xi  c_L - 6 \Delta  c_M - \xi \right)}{6 (2 \xi )^{3/2} \sqrt{c_M}}, 
\ee
and we have to choose as $(\sigma_c,\rho_c)$ whichever solution maximizes $f(\sigma,\rho)$.
We can see that
\bea
f(\sigma_0,\rho_0) &\approx& 2\pi\sqrt{\frac{\xi}{2c_M}}\left(c_L-\frac{1}{6}\right)+2\pi \Delta \sqrt{\frac{c_M}{2\xi}}, \cr
f(\tilde{\sigma}_0,\tilde{\rho}_0) &\approx& -2\pi\sqrt{\frac{\xi}{2c_M}}\left(c_L-\frac{1}{6}\right)-2\pi \Delta \sqrt{\frac{c_M}{2\xi}}. 
\eea
When taking $\Delta$ and $\xi$ to large values, we assume that, we do it in such a way that
\be
\lim_{\Delta,\xi\rightarrow \infty} \frac{\Delta}{\xi} = \gamma, 
\ee
where $\gamma$ is a finite number.
Then we have
\begin{subequations}
\bea
f(\sigma_0,\rho_0) &\approx& 2\pi\sqrt{\frac{\xi}{2c_M}} \left(c_L-\frac{1}{6} + \gamma c_M\right), \\
f(\tilde{\sigma_0},\tilde{\rho}_0) &\approx& -2\pi\sqrt{\frac{\xi}{2c_M}} \left(c_L-\frac{1}{6} + \gamma c_M\right).
\eea
\end{subequations}
$(\sigma_0,\rho_0)$ will be the maxima if $f(\sigma_0,\rho_0)>f(\tilde{\sigma_0},\tilde{\rho}_0)$. From the above equation, we can see that this will be the case when
\be
c_L-\frac{1}{6} + \gamma c_M > 0.
\ee
Likewise, $(\tilde{\sigma_0},\tilde{\rho}_0)$ is the maxima when
\be
-c_L + \frac{1}{6} - \gamma c_M  > 0 .  
\ee
Now, the saddle point approximation \eqref{density_saddle} is valid only when $\tilde{Z}\left(-\frac{1}{\sigma_c},\frac{\rho_c}{\sigma_c^2}\right)$ is dominated by the vacuum. We will separately analyse what conditions we have to impose for this to happen for the two cases mentioned above. 

\subsection*{Case 1: $(\sigma_0,\rho_0)$ is the maximum}
The arguments of the function $Z(-\frac{1}{\sigma_0},\frac{\rho_0}{\sigma_0^2})$ are 
\bea
-\frac{1}{\sigma_0} = i\sqrt{\frac{2\xi}{c_M}},\quad\quad \frac{\rho_0}{\sigma_0^2} = i\frac{1}{c_M}\sqrt{\frac{\xi}{2}} \left[\frac{1}{6}-c_L + \gamma c_M\right] \equiv i\frac{1}{c_M}\sqrt{\frac{\xi}{2}} \lambda.
\eea
From \eqref{partition_tilde} and \eqref{partition_function}, we have
\be
\tilde{Z}\left(-\frac{1}{\sigma_0},\frac{\rho_0}{\sigma_0^2}\right) = \tilde{D}(0,0)\left(1-e^{-2\pi \sqrt{\frac{2\xi}{c_M}}}\right)^2 + \sum_{(\Delta^{\prime},\xi^{\prime})\neq (0,0)} D(\Delta^{\prime},\xi^{\prime})\,e^{-2\pi \sqrt{\frac{2\xi}{c_M}}\Delta^{\prime}}e^{-2\pi \frac{1}{c_M}\sqrt{\frac{\xi}{2}} \lambda \xi^{\prime}}.
\ee
We assume that in our theory $L_0$ is bounded from below and hence $\Delta$'s takes only positive values. So, for large $\xi$
\be
e^{-2\pi \sqrt{\frac{2\xi}{c_M}}\Delta^{\prime}} \approx 0, 
\ee 
and we don't want $e^{-2\pi \frac{1}{c_M}\sqrt{\frac{\xi}{2}} \lambda \xi^{\prime}}$ to diverge in this limit. This means that $\lambda$ has to be either a finite positive number or zero. In other words
\be 
\lambda = \frac{1}{6}-c_L + \gamma \, c_M \geq 0.
\ee
If this is satisfied, then $\tilde{Z}(-\frac{1}{\sigma_0},\frac{\rho_0}{\sigma_0^2})$ is dominated by the contribution from the vacuum
\be
\tilde{Z}\left(-\frac{1}{\sigma_0},\frac{\rho_0}{\sigma_0^2}\right) \approx \tilde{D}(0,0).
\ee

\subsection*{Case 2: $(\tilde{\sigma}_0,\tilde{\rho}_0)$ is the maximum}
Now let us look at the conditions we have to impose so that $Z(-\frac{1}{\tilde{\sigma}_0},\frac{\tilde{\rho}_0}{\sigma_0^2})$ is dominated by the vacuum. The arguments of the function are 
\bea
-\frac{1}{\tilde{\sigma}_0} = -i\sqrt{\frac{2\xi}{c_M}},\quad\quad \frac{\tilde{\rho}_0}{\tilde{\sigma}_0^2} = i\frac{1}{c_M}\sqrt{\frac{\xi}{2}} \left[-\frac{1}{6}+c_L - \gamma c_M\right] \equiv i\frac{1}{c_M}\sqrt{\frac{\xi}{2}} \tilde{\lambda}.
\eea
And from \eqref{partition_tilde} and \eqref{partition_function}, we have
\be
\tilde{Z}\left(-\frac{1}{\tilde{\sigma}_0},\frac{\tilde{\rho}_0}{\tilde{\sigma}_0^2}\right) = \tilde{D}(0,0)\left(1-e^{2\pi \sqrt{\frac{2\xi}{c_M}}}\right)^2 + \sum_{(\Delta^{\prime},\xi^{\prime})\neq (0,0)} D(\Delta^{\prime},\xi^{\prime})\,e^{2\pi \sqrt{\frac{2\xi}{c_M}}\Delta^{\prime}}e^{-2\pi \frac{1}{c_M}\sqrt{\frac{\xi}{2}} \tilde{\lambda} \xi^{\prime}}.
\ee
Let us look at the second term. We can see that for large $\xi$, the factor $e^{2\pi \sqrt{\frac{2\xi}{c_M}}\Delta^{\prime}}$ diverges. So, we need at least the other factor, $e^{-2\pi \frac{1}{c_M}\sqrt{\frac{\xi}{2}} \lambda \xi^{\prime}}$, to vanish in this limit. In other words, $\tilde{\lambda}$ has to be a positive number
\be
-\frac{1}{6}+c_L - \gamma c_M >0. 
\ee
We also have 
\be
-c_L +\frac{1}{6}- \gamma c_M >0. 
\ee
The above two inequalities does not have any common overlap. So, we can conclude that $Z(-\frac{1}{\tilde{\sigma}_0},\frac{\tilde{\rho}_0}{\sigma_0^2})$ cannot be dominated by the vacuum and thus the saddle point analysis is not useful for this case.

\medskip 

\noindent We conclude that the density of primaries with large $\Delta$ and $\xi$ satisfying $c_L-\frac{1}{6} + \gamma c_M > 0$ and $\frac{1}{6} - c_L + \gamma c_M > 0$, using \eqref{density_saddle},  is given by
\be
D(\Delta,\xi) \approx \tilde{D}(0,0)\exp\left(2\pi\sqrt{\frac{\xi}{2c_M}}\left(c_L-\frac{1}{6}\right)+2\pi \Delta \sqrt{\frac{c_M}{2\xi}}+\frac{1}{2}\log\left(\frac{2\xi}{c_M}\right)\right).\label{Dprim}
\ee
Notice that the leading piece of the entropy, which is obtained by taking a logarithm of the density of states, obtained from \refb{Dprim} equals
\be
S_{\mbox{\tiny{Primary}}} = 2\pi\left(\sqrt{\frac{\xi}{2c_M}}\left(c_L-\frac{1}{6}\right)+\Delta \sqrt{\frac{c_M}{2\xi}}\right).
\ee
This is the same as \refb{bmscardy} with a replacement $c_L \to c_L - 1/6$. The entropy obtained from primaries is clearly the principle part of the whole entropy and it is clear that these contribute the largest when one is looking at holographic applications. 

To make contact with gravity in asymptotically flatspace, we recall that in 3d flat Einstein gravity $c_L=0$. The overlap of the inequalities thus reads 
$$ \gamma c_M > \frac{1}{6}.$$
This is a perfectly acceptable range of values and the saddle-point analysis works well in this regime. Also, the holographic regime is given by the limit of large central charge, which in this context means large $c_M$. The contribution to the correction piece $$\tilde{S}= S_{\mbox{\tiny{total}}} -  S_{\mbox{\tiny{Primary}}}= - \frac{\pi}{3} \sqrt{\frac{\xi}{2c_M}}$$ is clearly subleading in this limit. So the density of these BMS primaries captures the entropy of the flat space cosmologies rather well. 

\bigskip

\section{Conclusions}
\subsection*{Summary}
To summarise, we constructed the characters for the highest weight representations of the BMS$_3$ algebra in two different ways at the beginning of this paper. Some of the details of the proof using the BMS Gram matrix are detailed in Appendices \ref{appendix_1} and \ref{appendix_2}. 
\medskip

\noindent We then used two singular limits from 2d CFTs to reproduce our character formula from the well-known Virasoro characters. This led to a conundrum, since one of the limits (the non-relativistic one) took Virasoro highest representations to BMS highest weight representations, and the other (the ultra-relativistic one) took the  Virasoro highest representations to the very different BMS induced representation. We attempted to explain this by alluding to a novel automorphism in the parent 2d CFT. 
\medskip

\noindent Finally, we used the form of the characters to construct the partition function and from there derived the density of BMS primary states in the spectrum. We were able to see, interestingly, the BMS primaries capture the principle part of the BMS-Cardy entropy, of e.g. the FSC solutions.

\subsection*{Future directions}
As we have already emphasised, we wish to investigate this novel automorphism in the context of 2d CFTs. This, to the best of our knowledge, is the first time this rather peculiar automorphism has been noticed in literature {\footnote{A supersymmetric version of this automorphism has been reported in \cite{Bagchi:2018wsn} in relation to tensionless strings, but  observation was made in the context of this work earlier.}}. This should lead to a deeper understanding of the rather bizarre duality between non-relativistic and ultra-relativistic physics, and also should go a long way to understand the representation theory of the BMS$_3$ algebra. This should also lead to a clarification of which particular representation to use while constructing the field theory dual to 3d asymptotically flat space. The highest weight representations are very clearly more useful for computational purposes. It seems that because of this isomorphism, some answers we get are independent of the underlying representation. We wish to understand in detail to what extent this feature is true.

\medskip

\noindent We investigate various structure related to the BMS Gram matrix in the two appendices. A very natural question is to investigate the equivalent of the Kac determinant for the BMS$_3$ algebra. We should be able to relate this to the null vectors of the BMS module investigated e.g. in \cite{Bagchi:2009pe}. This, unfortunately, does not seem as straightforward as it may seem. Our constructions so far only reproduce the most ``singular" null vector (constructed out of $M_n$'s alone) for an arbitrary level. We are looking to solve this issue and understand whether there is a minimal series for the BMS$_3$ algebra like the Virasoro. In this context, it is interesting to point out that in \cite{Hijano:2018nhq} while generalising the monodromy method of calculating BMS blocks (beyond the LLLL or Poincare blocks as computed in \cite{Bagchi:2016geg, Bagchi:2017cpu} and holographically in \cite{Hijano:2017eii}), some rather strange representations, which are reducible but indecomposable like logarithmic CFTs, were used. It may be of interest to reconsider these in the context of the BMS Kac determinant and matching of null states.

\medskip

\noindent There are of course rather straight-forward generalisations of this work that we are interested in pursuing. We would like to find the characters of higher-spin versions of BMS \cite{Afshar:2013vka, Gonzalez:2013oaa} (called BMW algebras!), and different supersymmetric versions of BMS$_3$ (e.g. the ``homogeneous" \cite{Barnich:2014cwa, Barnich:2015sca, Bagchi:2016yyf, Lodato:2016alv, Lodato:2018gyp} and the ``inhomogeneous" \cite{Lodato:2016alv, Bagchi:2016yyf, Bagchi:2017cte, Lodato:2018gyp} algebras). Some of the above (the spin-3 case and the ``homogeneous" superalgebra) have been computed for the induced representations in \cite{Campoleoni:2015qrh}. It is of interest to see how much of their analysis is reproduced in the highest weight analysis. It would be even more interesting to see if there are any departures from the answers of the induced representations. One expects to see some differences in the higher spin versions. 

\medskip

\noindent On a gravitational side, the one loop determinant of the bulk theory equals the character of the boundary symmetry \cite{Maloney:2007ud, Giombi:2008vd, David:2009xg}. This has been computed also for flat-spacetime in \cite{Barnich:2015mui}, where the answer is exactly the character formula we have computed here. A generalisation of this analysis is the computation of one-loop determinants for higher spin theories in the asymptotically flat 3d bulk and also the 3d supergravity theory which has the homogeneous super-BMS algebras as its asymptotic symmetries \cite{Barnich:2014cwa, Barnich:2015sca}. This has been performed in \cite{Campoleoni:2015qrh}. We would like to investigate the inhomogeneous super-algebra \cite{Lodato:2016alv, Bagchi:2016yyf, Bagchi:2017cte} arising out of the ``twisted" supergravity construction of  \cite{Lodato:2016alv}. 

\medskip

\noindent One of the more ambitious programmes is to attempt a microscopic counting of states of the Flatspace Cosmologies, {\em{{\'a} la}} Strominger-Vafa \cite{Strominger:1996sh}. The fact that the character of the BMS module has a positive integral expansion in terms of the levels and clearly counts the number of states of the putative dual field theory, gives us hope of a more fundamental string theoretic understanding of the underlying degrees of freedom. A string/M theory embedding of the FSC was initially discussed in \cite{Cornalba:2002fi}. We wish to construct the equivalent of the D1-D5 CFT on the field theory side, and one of the first things to attempt is a contraction of the superalgebra and this should lead to a better understanding of the symmetries on the field theoretic side. It should turn out to be one of the different $\mathcal{N}=4$ Super-BMS algebras that can be constructed in close analogy with the homogeneous and inhomogeneous algebras discussed above. The systematic limit could also be a way to understand the brane construction analogue to the D1-D5 system.

\bigskip

\section*{Acknowledgements}
Discussions and correspondence with Rudranil Basu, Diptarka Das, Abhijit Gadde, Daniel Grumiller, and Gautam Mandal are gratefully acknowledged. This work was presented in TIFR Mumbai, SINP Kolkata and Durham University prior to publication. AB thanks the string theory groups at these places for interesting discussions.  

\medskip

\noindent AB also acknowledges the warm hospitality of the Department of Mathematics and Grey College at Durham University and the financial support of the Institute of Advanced Study, Durham through a Senior Research Fellowship (COFUNDed between Durham University and the European Union under grant agreement number 609412), during the final stages of this work.

\medskip

\noindent This work is also partially supported by the following grants: DST-INSPIRE faculty award, SERB Early Career Research Award (ECR/2017/000873), DST-Max Planck mobility award, DST-BMWF India-Austria bilateral grant, Royal Society International Exchange grant, SERB extra-mural grant (EMR/2016/008037), SERB National Post Doctoral Fellowship PDF/2016/002166.

\bigskip

\appendix
\section*{APPENDICES}
\section{Structure of the BMS Gram matrix \label{appendix_1}}

In this appendix we will show that the Gram matrix can always be put in the following form if we order the basis states according to some rules which we will lay out.
\be
K^{(N=\text{odd})} = \left(\begin{array}{ccccc}
K_{1,1} &  & \cdots &  & A_{\widetilde{\rm dim}_N}\\
 &  &  & \iddots & 0\\
\vdots &  & \iddots &  & \vdots\\
 & A_{2}\\
A_{1} & 0 & \cdots &  & 0
\end{array}\right) ,
\label{oddmatrix}
\ee
\be
K^{(N=\text{even})}=\left(\begin{array}{cccccccccc}
K_{1,1} & \cdots & \cdots &  & \cdots &  & \cdots &  & \cdots & A_{\widetilde{{\rm dim}}_{N}}\\
\vdots & \ddots &  &  &  &  &  &  & \iddots & 0\\
 &  & \ddots &  &  &  &  & A_{k+l} &  & \vdots\\
 &  &  & D_{1} & 0 & \cdots & 0\\
\vdots &  &  & 0 & D_{2} & \cdots & 0 &  &  & \vdots\\
\vdots &  &  & \vdots & \vdots & \ddots & \vdots &  &  & \vdots\\
 &  &  & 0 & 0 & \cdots & D_{l}\\
 &  & A_{k} &  &  &  &  & 0\\
\vdots & \iddots &  &  &  &  &  &  & \ddots & \vdots\\
A_{1} & 0 & \cdots &  & \cdots &  & \cdots &  & \cdots & 0
\end{array}\right).
\label{evenmatrix}
\ee
We can see from the above figure that the difference between the Gram matrix for odd and even level is that for even level we have a diagonal matrix at the matrix. The dimension of this diagonal matrix is $p(N/2)$. With these structure for $K^{(N)}$, the inverse matrix $K_{(N)}$ will have the form
\be
K_{(N=\text{odd})} = \left(\begin{array}{ccccc}
0 & \cdots & 0 & \cdots & \frac{1}{A_{\widetilde{\rm dim}_N}}\\
0 &  &  & \iddots & \vdots\\
\vdots &  & \iddots &  & \vdots\\
 & \frac{1}{A_{2}}\\
\frac{1}{A_{1}} & \iddots & \cdots &  & K^{\widetilde{\rm dim}_N,\widetilde{\rm dim}_N}
\end{array}\right) ,
\label{oddmatrix_in}
\ee
\be
K_{(N=\text{even})}=\left(\begin{array}{cccccccccc}
0 & \cdots & \cdots &  & \cdots &  & 0 &  & \cdots & \frac{1}{A_{\widetilde{{\rm dim}}_{N}}}\\
\vdots & \ddots &  &  &  &  &  &  & \iddots & \\
 &  & \ddots &  &  &  &  & \frac{1}{A_{k+l}} &  & \vdots\\
 &  &  & \frac{1}{D_{1}} & 0 & \cdots & 0\\
\vdots &  &  & 0 & \frac{1}{D_{2}} & \cdots & 0 &  &  & \vdots\\
0 &  &  & \vdots & \vdots & \ddots & \vdots &  &  & \vdots\\
 &  &  & 0 & 0 & \cdots & \frac{1}{D_{l}}\\
 &  & \frac{1}{A_{k}} &  &  &  &  & \ddots \\
\vdots & \iddots &  &  &  &  &  &  & \ddots & \vdots\\
\frac{1}{A_{1}} &  & \cdots &  & \cdots &  & \cdots &  & \cdots & K^{\widetilde{\rm dim}_N,\widetilde{\rm dim}_N}
\end{array}\right).
\label{evenmatrix_in}
\ee
\medskip 
\subsection{Conditions for Non-zero BMS Inner Products}
In order to find the ordering rules, we first have to find the conditions for inner products of states in the BMS module to be non-zero.
In the usual eigenbasis of $L_0$, the linearly independent descendant states of a primary state $|\Delta, \xi \rangle$ are chosen to be $$L_{-i_1}^{l_1}L_{-i_2}^{l_2} \ldots L_{-i_r}^{l_r}M_{-j_1}^{m_1}M_{-j_2}^{m_2} \ldots M_{-j_s}^{m_s}|\Delta, \xi \rangle,$$ where $1\leqslant i_1 < i_2 <...< i_r$ and $1\leqslant j_1 < j_2 <...< j_s$. This is a descendant state at level $N= \sum_{p=1}^r i_p l_p + \sum_{q=1}^s j_q m_q $.
In this basis, we can readily check the conditions for non-vanishing inner product between the basis vectors. Let us consider two generic states: $$L_{-i_1}^{l_1}L_{-i_2}^{l_2}...L_{-i_r}^{l_r}M_{-j_1}^{m_1}M_{-j_2}^{m_2}...M_{-j_s}^{m_s}|\Delta, \xi \rangle \quad \& \quad L_{-i'_1}^{l'_1}L_{-i'_2}^{l'_2}...L_{-i'_{r'}}^{l'_{r'}}M_{-j'_1}^{m'_1}M_{-j'_2}^{m'_2}...M_{-j'_{s'}}^{m'_{s'}}|\Delta, \xi \rangle.$$ There are two basic conditions that are required to be satisfied for non-zero inner products between these two states:
\begin{enumerate}[leftmargin=*]
\item These two states must be at the same level
\be
 \sum_{p=1}^r i_p l_p + \sum_{q=1}^s j_q m_q = \sum_{p=1}^{r'} i'_p l'_p + \sum_{q=1}^{s'} j'_q m'_q.
\ee

\item Inside the bra-ket of inner product, i.e., in the expression 
\be 
\langle \Delta, \xi|M_{j'_{s'}}^{m'_{s'}}...M_{j'_2}^{m'_2}M_{j'_1}^{m'_1}L_{i'_{r'}}^{l'_{r'}}...L_{i'_2}^{l'_2}L_{i'_1}^{l'_1}L_{-i_1}^{l_1}L_{-i_2}^{l_2}...L_{-i_r}^{l_r}M_{-j_1}^{m_1}M_{-j_2}^{m_2}...M_{-j_s}^{m_s}|\Delta, \xi \rangle ,
\ee
for every $M_{j'_{q}}$  ($1\leqslant q \leqslant s'$) we must have combinations of L-operators that give (after using the commutation relations or just simply summing the L-indices) $L_{-j'_{q}}$.  Similarly, for every $M_{-j_{q}}$  we must have combinations of L-operators that gives $L_{j_{q}}$ . Note that if we have more than one $M_{j'_{q}}$  we need separate $L_{-j'_{q}}$ for each of them. The same thing applies for $M_{-j_{q}}$. The combinations (or sets) of L-operators once used, are not further considered (all the elements of those sets) to find suitable combinations for remaining M-operators.
\end{enumerate}

From the above two basic rules and working out some examples, we can derive some easy-to-work-with conditions for non-zero inner product. Before we give these rules, let us first introduce some notations and definitions which will makes things simpler. We associate two numbers $\alpha$ and $\beta$ for each basis states $L_{-i_1}^{l_1}L_{-i_2}^{l_2}...L_{-i_r}^{l_r}M_{-j_1}^{m_1}M_{-j_2}^{m_2}...M_{-j_s}^{m_s}|\Delta, \xi \rangle$ which are given by
\bea
\alpha &=& \text{No. of L-string - No. of M-string} = \sum_{p=1}^r l_p -\sum_{q=1}^s m_q, \cr
\beta  &=& \text{Sum of L-indices - Sum of M-indices} = \sum_{p=1}^r i_pl_p -\sum_{q=1}^s j_p m_q. 
\eea
In particular, at level $N$, the value of $\alpha$ range from $N$ (associated with the basis state $L_{-1}^N|\Delta,\xi\rangle$) to $-N$ (associated with the state $M_{-1}^N|\Delta,\xi\rangle$). There will be symmetry in the sense that the number of states sharing the same value of $(\alpha,\beta)$ will be equal to the number of states with same value of $(-\alpha,-\beta)$. For example at level 5, there are two states $(L_{-2}^2M_{-1}|\Delta,\xi\rangle,\,L_{-1}L_{-3}M_{-1}|\Delta,\xi\rangle)$ with $(\alpha=1,\beta=3)$. Similarly there are two states $(L_{-1}M_{-2}^2|\Delta,\xi\rangle,\,L_{-1}M_{-1}M_{-3}M_{-1}|\Delta,\xi\rangle)$ with $(\alpha=-1,\beta=-3)$. It can be seen that these two sets of states are related by swapping $L$ and $M$. We will call this operation conjugation
\be
\text{Conjugation},\,\,\,L \longleftrightarrow M,
\ee
and pairs like $L_{-2}^2M_{-1}|\Delta,\xi\rangle$ and $L_{-1}M_{-2}^2|\Delta,\xi\rangle$ which are related through conjugation to be a conjugate pair. So, conjugation flips the sign of $\alpha$ and $\beta$. We could also have self conjugate states i.e., states which remain the same under conjugation. The simplest example is $L_{-1}M_{-1}|\Delta,\xi\rangle$. These kind of states only exists for even level and have $(\alpha=0,\beta=0)$. Using these notations, the non-zero conditions for the inner product of basis states $|\Psi\rangle$ and $|\Psi^{\prime}\rangle$ are:
\begin{enumerate}
\item Inside the bra-ket of inner product, the total number of L-operators must not be less than the total number of M-operators. Otherwise, the inner product is zero
\be
\langle \Psi|\Psi^{\prime}\rangle = 0, \,\,\text{if}\,\,\alpha+\alpha^{\prime}<0.
\label{rule1}
\ee 
\item If the total number of L-operators is greater than that of M, the sum of L-indices must not be less than the sum of M-indices
\bea
\text{If} && \alpha+\alpha^{\prime}>0, \,\,
\text{we require} \,\, \beta+\beta^{\prime}>0 .
\eea
\item The next rule is for the case when the total number of L-operators and M-operators are the same i.e., when $\alpha+\alpha^{\prime}=0$
\bea
{\rm If} \,\,\alpha + \alpha^{\prime}=0,\,\,{\rm we\,require}&& 
 i)\,\,r=s',\,\,r'=s, \cr
&& ii)\,\, i_p=j'_p,\, l_p=m'_p,\,\,{\rm for} \,1\leqslant p \leqslant r,  \cr
&& iii)\,\,i'_q=j_q,\,\,l'_q=m_q,\,\,{\rm for} 1\leqslant q \leqslant s .
\label{rule3}
\eea
This means that whenever the total number of L-operators and M-operators are the same, the two states must be a conjugate pair.
\end{enumerate}
All other pairs of states not obeying the above rules give vanishing inner product.
 Using the above rules,  we can readily deduce the conditions (not sufficient but necessary) for a generic state to have non-zero norms. These are
\bea 
&& \alpha \geq 0 , \\
&& \text{If}\,\alpha>0,\,\,\text{then we require}\,\,\beta>0,\, 
\\
&& \text{If}\,\,\alpha= 0,\,\,\text{the state should be self conjugate}.
\eea
We have already mentioned that at odd level we could not have self conjugate states. So, at odd levels, all the states with equal length of L-string and M-string i.e., $\alpha=0$ have vanishing norm.

\subsection{The ordering rules}

At a particular level of a generic BMS module, we can now find the pairs of basis states that have vanishing inner product using the above rules.  From this we can guess a method of arranging the basis states such that the resulting Gram matrix has a structure mentioned in the beginning of this appendix. These arrangement rules are given by:
\begin{enumerate} [leftmargin=*]
\item Arrange the basis states in decreasing order of $\alpha$, keeping $L_{-1}^n|\Delta, \xi \rangle $ at the beginning and $M_{-1}^n|\Delta, \xi \rangle $ at the ending of the queue. Keep the states having the same $\alpha$ one after another as a bunch.

\item For states with the same value of $\alpha$, ordering is done in decreasing value of $\beta$.

\item If we have $r$ states with the same value of $(\alpha,\beta)\neq (0,0)$ and arrange them in the order $(|\Phi_1\rangle,\,\,|\Phi_2\rangle,....,|\Phi_{r-1}\rangle,\,\,|\Phi_r\rangle)$, then the $r$ conjugate of these states with $(-\alpha,-\beta)$ must be arranged in the order $(|\Phi_r^c\rangle,\,\,|\Phi_{r-1}^c\rangle,....,|\Phi_2^c\rangle,\,\,|\Phi_1^c\rangle)$, where $c$ means a conjugation. In other words, if a state is at position $n$, then its conjugate state should be at the position $\widetilde{\dim}_N-n+1$, where $\widetilde{\dim}_N$ is the number of states at level N. So, every state will pair up with its conjugate state at the anti-diagonal line of the Gram matrix. 

\item For even level $N$, we could have $p(N/2)^2$ states with $(\alpha=0,\beta=0)$ and $p(N/2)$ of these are self-conjugate states. We first put the self-conjugate states (in any order) at the centre and arrange the other states in such way that if a states is at position $s$, somewhere on the left side of the centre, its conjugate should be on the right side of the centre at position  $p(N/2)^2-s+1$. This arrangement is similar to the preceding rule where the aim is again to let conjugate pairs meet at the anti-diagonal line in the Gram matrix.

\end{enumerate}

We have already seen that the above ordering rules for level 1 and level 2
give us a triangular Gram matrix in \eqref{GM_level1} and \eqref{GM_level2}.
For level 3, the ordering of the states are given in Table \ref{BMS_states}. We shown the states again below along with their values of $\alpha$ and $\beta$ 
\bea
&&
\left(\begin{array}{l|llll}
 & \beta=3 & \beta=1 & \beta=-1 & \beta=-3\\
\hline \alpha=3 & |\Psi_{1}\rangle=L_{-1}^{3}|\Delta,\xi\rangle\\
\alpha=2 & |\Psi_{2}\rangle=L_{-1}L_{-2}|\Delta,\xi &  &  & \\
\alpha=1 & |\Psi_{3}\rangle=L_{-3}|\Delta,\xi\rangle & |\Psi_{4}\rangle=L_{-1}^{2}M_{-1}|\Delta,\xi\rangle\\
\alpha=0 &  & |\Psi_{5}\rangle=L_{-2}M_{-1}|\Delta,\xi\rangle & |\Psi_{6}\rangle=L_{-1}M_{-2}|\Delta,\xi\rangle\\
\alpha=-1 &  &  & |\Psi_{7}\rangle=L_{-1}M_{-1}^{2}|\Delta,\xi\rangle & |\Psi_{8}\rangle=M_{-3}|\Delta,\xi\rangle\\
\alpha=-2 &  &  &  & |\Psi_{9}\rangle=M_{-1}M_{-2}|\Delta,\xi\rangle\\
\alpha=-3 &  &  &  & |\Psi_{10}\rangle=M_{-1}^{3}|\Delta,\xi\rangle
\end{array}\right).\cr
&&
\eea
The form of the Gram matrix with this ordering ordering of basis is indeed the one we expected 
\be
\left(\begin{array}{c|cccccccccc}
 & |\Psi_{1}\rangle & |\Psi_{2}\rangle & |\Psi_{3}\rangle & |\Psi_{4}\rangle & |\Psi_{5}\rangle & |\Psi_{6}\rangle & |\Psi_{7}\rangle & |\Psi_{8}\rangle & |\Psi_{9}\rangle & |\Psi_{10}\rangle\\
\hline \langle\Psi_{1}| & K_{1,1} & K_{1,2} & K_{1,3} & K_{1,4} & K_{1,5} & K_{1,6} & K_{1,7} & K_{1,8} & K_{1,9} & K_{1,10}\\
\langle\Psi_{2}| & K_{2,1} & K_{2,2} & K_{2,3} & K_{2,4} & K_{2,5} & K_{2,6} & K_{2,7} & K_{2,8} & K_{2,9} & 0\\
\langle\Psi_{3}| & K_{3,1} & K_{3,2} & K_{3,3} & K_{3,4} & K_{3,5} & K_{3,6} & 0 & K_{3,8} & 0 & 0\\
\langle\Psi_{4}| & K_{4,1} & K_{4,2} & K_{4,3} & K_{4,4} & K_{4,5} & K_{4,6} & K_{4,7} & 0 & 0 & 0\\
\langle\Psi_{5}| & K_{5,1} & K_{5,2} & K_{5,3} & K_{5,4} & 0 & K_{5,6} & 0 & 0 & 0 & 0\\
\langle\Psi_{6}| & K_{6,1} & K_{6,2} & K_{6,3} & K_{6,4} & K_{6,5} & 0 & 0 & 0 & 0 & 0\\
\langle\Psi_{7}| & K_{7,1} & K_{7,2} & 0 & K_{7,4} & 0 & 0 & 0 & 0 & 0 & 0\\
\langle\Psi_{8}| & K_{8,1} & K_{8,2} & K_{8,3} & 0 & 0 & 0 & 0 & 0 & 0 & 0\\
\langle\Psi_{9}| & K_{9,1} & K_{9,2} & 0 & 0 & 0 & 0 & 0 & 0 & 0 & 0\\
\langle\Psi_{10}| & K_{10,1} & 0 & 0 & 0 & 0 & 0 & 0 & 0 & 0 & 0
\end{array}\right).
\ee
For level 4, a possible ordering of the basis states using our arrangement rules is given by
\bea 
&& |\Psi_1\rangle = L_{-1}^4|\Delta,\xi\rangle,\,\,\,|\Psi_2\rangle = L_{-1}^2L_{-2}|\Delta,\xi\rangle,\,\,\,|\Psi_3\rangle = L_{-2}^2|\Delta,\xi\rangle,\,\,\, |\Psi_4\rangle = L_{-1}L_{-3}|\Delta,\xi\rangle, \cr
&& |\Psi_5\rangle = L_{-1}^3M_{-1}|\Delta,\xi\rangle,\,\,\, |\Psi_6\rangle = L_{-4}|\Delta,\xi\rangle,\,\,\,|\Psi_7\rangle = L_{-1}L_{-2}M_{-1}|\Delta,\xi\rangle,\,\,\,|\Psi_8\rangle = L_{-1}^2M_{-2}|\Delta,\xi\rangle,\cr
&& |\Psi_9\rangle = L_{-3}M_{-1}|\Delta,\xi\rangle,\,\,\,|\Psi_{10}\rangle = L_{-2}M_{-2}|\Delta,\xi\rangle,\,\,\,|\Psi_{11}\rangle = L_{-1}^2M_{-1}^2|\Delta,\xi\rangle,\,\,\,|\Psi_{12}\rangle = L_{-1}M_{-3}|\Delta,\xi\rangle\cr
&& |\Psi_{13}\rangle = L_{-2}M_{-1}^2|\Delta,\xi\rangle ,\,\,\, |\Psi_{14}\rangle = L_{-1}M_{-1}M_{-2}|\Delta,\xi\rangle,\,\,\, |\Psi_{15}\rangle = M_{-4}|\Delta,\xi\rangle,\,\,\,|\Psi_{16}\rangle = L_{-1}M_{-1}^3|\Delta,\xi\rangle \cr
&& |\Psi_{17}\rangle = M_{-1}M_{-3}|\Delta,\xi\rangle,\,\,\,|\Psi_{18}\rangle = M_{-2}^2|\Delta,\xi\rangle ,\,\,\, |\Psi_{19}\rangle = M_{-1}^2M_{-2}|\Delta,\xi\rangle ,\,\,\, |\Psi_{20}\rangle = M_{-1}^4|\Delta,\xi\rangle.\cr
&&
\eea

Now let us explain why the above rules give us a triangular structure for the Gram matrix. We first consider the case of odd level. Here we don't have to deal with self-conjugate states and our arrangement rules is such that the matrix entries of the anti-diagonal line are the inner products of a state and its conjugate and is thus non-zero due to \eqref{rule3}. Let us look at a particular anti-diagonal element pairing up a state $|\Psi_a\rangle$ with its conjugate $|\Psi_b\rangle$
\be
K_{a,b}=\langle \Psi_a | \Psi_b \rangle \neq 0, \,\,\alpha_a+\alpha_b=0, \,\,\beta_a+\beta_b=0. 
\ee
Any matrix elements on its right side is given by
\be
K_{a,b+c} = \langle \Psi_a | \Psi_{b+c} \rangle.
\ee
Since we are arranging the states in such a that way $\alpha_b\geq\alpha_{b+1}$, we could have $\alpha_b=\alpha_{b+c}$ or $\alpha_b>\alpha_{b+c}$. For the first case we have 
\be
\alpha_a + \alpha_{b+c} = 0. 
\ee
From \eqref{rule3}, for the inner product $\langle \Psi_a|\Psi_{b+c}\rangle$ to be non-zero, $|\Psi_{b+c}\rangle$ should be a conjugate of $|\Psi_a\rangle$. But this is not possible as $|\Psi_b\rangle$ is the conjugate of $|\Psi_a\rangle$. Therefore
\be
\langle \Psi_a | \Psi_{b+c} \rangle=0. 
\ee
For the next case, from \eqref{rule1}
\be
 \langle\Psi_{a}|\Psi_{b+c}\rangle = 0, \,\,\text{as}\,\,\alpha_a+\alpha_{b+c}<0.
\ee
So all matrix entries on the right side of $\langle \Psi_a|\Psi_b\rangle$ are zero. In other words, the Gram matrix is triangular. 

For even level $N$, we have $p(N/2)$ self-conjugate states. These have non-zero norms and are orthogonal to each other due to \eqref{rule3}. So, the Gram matrix constructed from these sub-set of states will be diagonal and this diagonal matrix will sit at the centre of the Gram matrix due to our arrangement rules of putting self-conjugate states at the middle while ordering the basis states. Similar arguments made for odd level can be used to conclude that all matrix elements on the right hand side of this diagonal matrix will be zero. Just like odd level, all the non self-conjugate states pair up with their conjugate at the anti-diagonal line outside the diagonal matrix, with all the entries on the right side of this anti-diagonal line vanishing again due to the same argument. Thus we have proven that with the use of our arrangement rules the Gram matrix will have the form \eqref{oddmatrix} for odd level and the form \eqref{evenmatrix} for even level.

\section{Form of $\sum_{i,j}K^{ij}_{(N)}\tilde{O}_{ij}^{(N)}$ for general level $N$ \label{appendix_2}}
In this appendix we will prove the formula for $\sum_{i,j}K^{ij}_{(N)}\tilde{O}_{ij}^{(N)}$ given in \eqref{trace_level_N}. For this, we have to prove that, using our ordering rules for the basis states, the matrix $\tilde{O}^{(N)}$ have the following structure:
\be
\tilde{O}^{(N=\text{odd})} = e^{-2\pi i(\sigma\frac{c_L}{2}+\rho\frac{c_M}{2})}e^{2\pi i\sigma(\Delta+N)}e^{2\pi i\rho\xi}\left(\begin{array}{ccccc}
\tilde{O}_{1,1}^{\prime} &  & \cdots &  & A_{\widetilde{\rm dim}_N}\\
 &  &  & \iddots & 0\\
\vdots &  & \iddots &  & \vdots\\
 & A_{2}\\
A_{1} & 0 & \cdots &  & 0
\end{array}\right) ,
\ee
\be
\tilde{O}^{(N=\text{even})} = e^{-2\pi i(\sigma\frac{c_L}{2}+\rho\frac{c_M}{2})}e^{2\pi i\sigma(\Delta+N)}e^{2\pi i\rho\xi}\left(\begin{array}{cccccccccc}
\tilde{O}_{1,1}^{\prime} & \cdots & \cdots &  & \cdots &  & \cdots &  & \cdots & A_{\widetilde{{\rm dim}}_{N}}\\
\vdots & \ddots &  &  &  &  &  &  & \iddots & 0\\
 &  & \ddots &  &  &  &  & A_{k+l} &  & \vdots\\
 &  &  & D_{1} & 0 & \cdots & 0\\
\vdots &  &  & 0 & D_{2} & \cdots & 0 &  &  & \vdots\\
\vdots &  &  & \vdots & \vdots & \ddots & \vdots &  &  & \vdots\\
 &  &  & 0 & 0 & \cdots & D_{l}\\
 &  & A_{k} &  &  &  &  & 0\\
\vdots & \iddots &  &  &  &  &  &  & \ddots & \vdots\\
A_{1} & 0 & \cdots &  & \cdots &  & \cdots &  & \cdots & 0
\end{array}\right).
\ee
In the above equation, the $A$'s and $D$'s are the same as those appearing in \eqref{oddmatrix} and \eqref{evenmatrix} for $K^{(N)}$ and $K_{(N)}$. Before we prove the structural property of $\tilde{O}^{(N)}$ given above, let us first see how it lead to the formula \eqref{trace_level_N}. For odd level the reasoning is exactly same as what we did for level 2. More precisely, $K_{(N)}$ and $\tilde{O}^{(N)}$ have the opposite triangular structure so that when taking the trace, only the anti-diagonal elements will contribute. The anti-diagonal elements of these matrix are also inverse of each other apart from a factor of $e^{-2\pi i(\sigma\frac{c_L}{2}+\rho\frac{c_M}{2})}e^{2\pi i\sigma(\Delta+N)}e^{2\pi i\rho\xi}$. So we have
\bea
\sum_{i,j}K^{ij}_{(N)}\tilde{O}_{ij}^{(N)} &=& \sum_{i=1}^{\widetilde{\rm Dim}_N} K^{ii}_{(N)}\tilde{O}_{ii}^{(N)} = e^{-2\pi i(\sigma \frac{c_L}{2}+\rho \frac{c_M}{2})} e^{2\pi i\sigma (\Delta+N)}e^{2\pi i\rho \xi} \sum_{d=1}^{\widetilde{\rm Dim}_N} 1\cr
&=& \widetilde{\rm Dim}_N\, e^{-2\pi i(\sigma \frac{c_L}{2}+\rho \frac{c_M}{2})} e^{2\pi i\sigma (\Delta+N)}e^{2\pi i\rho \xi}.
\eea
This reasoning can be easily extended for even level. 

Now let us give a proof why  $\tilde{O}^{(N)}$ have the structure shown above if we use our ordering rules for the basis states.
We know from \eqref{act_L0} that the factor $e^{-2\pi i(\sigma\frac{c_L}{2}+\rho\frac{c_M}{2}}e^{2\pi i\sigma(\Delta+N)}$ is due to $e^{2\pi i\sigma (L_{0}-c_L/2)}e^{-2\pi i\rho c_M/2)}$. So, what is left is to see the matrix element of $e^{2\pi i\rho M_{0}}$. First let us note that using the commutator
\bea
e^{2\pi i\rho M_{0}} L_{\vec{k}}M_{\vec{q}}|\Delta,\xi\rangle &=& L_{\vec{k}}M_{\vec{q}} e^{2\pi i\rho M_{0}} |\Delta,\xi\rangle + [e^{2\pi i\rho M_{0}}, L_{\vec{k}}M_{\vec{q}}] |\Delta,\xi\rangle \cr
& = & e^{2\pi i\rho \xi}  L_{\vec{k}}M_{\vec{q}}|\Delta,\xi\rangle + \sum_{n=1}^{\infty} \frac{1}{n!}(2\pi i \rho)^n [M_0^n, L_{\vec{k}}M_{\vec{q}}] |\Delta,\xi\rangle.
\label{act_M0_gen}
\eea
As an example, we have
\bea
[e^{2\pi i\rho M_{0}}, L_{-1}L_{-1}] |\Delta,\xi\rangle &=& e^{2\pi i\rho\xi}\left(4\pi i\rho L_{-1}M_{-1} + M_{-1}M_{-1}\right)|\D,\xi\>.
\eea
The point is that since the commutator of $M_0$ with $L$'s changes $L$'s to $M$'s, all the states in 
$[e^{2\pi i\rho M_{0}}, L_{\vec{k}}M_{\vec{q}}] |\Delta,\xi\rangle$ will have less number of $L$'s, i.e., less value of $\alpha$, than that of $L_{\vec{k}}M_{\vec{q}}|\Delta,\xi\rangle$. The exception is for states of the form $M_{\vec{q}}|\Delta,\xi\rangle$ which will not be relevant for our discussion below. 
Now let us look at a particular anti-diagonal element of $O^{(N)}$ for odd level which pair up a state $|\Psi_a\rangle$ with its conjugate $|\Psi_b\rangle=L_{\vec{q}}M_{\vec{k}}|\Delta,\xi\rangle$. Using \eqref{act_M0_gen}, we have
\bea
\tilde{O}^{(N)}_{ab} &=&   e^{-2\pi i(\sigma\frac{c_L}{2}+\rho\frac{c_M}{2})}e^{2\pi i\sigma(\Delta+N)} \langle \Psi_a |e^{2\pi i\rho M_{0}} L_{\vec{q}}M_{\vec{k}}|\Delta,\xi\rangle \cr
&=& e^{-2\pi i(\sigma\frac{c_L}{2}+\rho\frac{c_M}{2})}e^{2\pi i\sigma(\Delta+N)} \left(e^{2\pi i\rho \xi}  \langle\Psi_a |\Psi_b \rangle+ \langle\Psi_a|[e^{2\pi i\rho M_{0}}, L_{\vec{q}}M_{\vec{k}}] |\Delta,\xi\rangle\right).\cr
&&
\eea
Now since $|\Psi_b\rangle$ and $|\Psi_b\rangle$ are conjugate pair we have 
\be
\alpha_a + \alpha_b = 0. 
\ee
As we have argued before, all the states in $[e^{2\pi i\rho M_{0}}, L_{\vec{q}}M_{\vec{k}}] |\Delta,\xi\rangle$ will have less value of $\alpha$ than that of $\alpha_b$. So, for these states we will have
\be
\alpha_a + \alpha < 0. 
\ee 
Therefore the inner products of these states with $|\Psi_a\rangle$ is zero due to \eqref{rule1}. In other words,
\be
\langle\Psi_a|[e^{2\pi i\rho M_{0}}, L_{\vec{q}}M_{\vec{k}}] |\Delta,\xi\rangle = 0.
\ee
Thus we have shown that
\be
 \tilde{O}^{(N)}_{ab} =  e^{-2\pi i(\sigma\frac{c_L}{2}+\rho\frac{c_M}{2})}e^{2\pi i\sigma(\Delta+N)}e^{2\pi i\rho\xi} K^{(N)}_{ab}.
\ee 
for anti-diagonal elements. Now let us look at matrix entries on the right side of the anti-diagonal element $\tilde{O}_{ab}^{(N)}$. Particularly, let us look at $\tilde{O}_{a(b+c)}^{(N)}$ with $|\Psi_{b+c}\rangle = L_{\vec{q^{\prime}}}M_{\vec{k^{\prime}}}$.  Using $K_{a(b+c)}=\langle \Psi_a | \Psi_{b+c}\rangle = 0$, we have 
\bea
\tilde{O}^{(N)}_{a(b+c)} &=& e^{-2\pi i(\sigma\frac{c_L}{2}+\rho\frac{c_M}{2})}e^{2\pi i\sigma(\Delta+N)} \langle\Psi_a|[e^{2\pi i\rho M_{0}}, L_{\vec{q^{\prime}}}M_{\vec{k^{\prime}}}] |\Delta,\xi\rangle.
\eea 
Due to our ordering rule we have $\alpha_{b+c}\leq\alpha_b$. We can again use the previous argument to conclude that all the states in $[e^{2\pi i\rho M_{0}}, L_{\vec{q^{\prime}}}M_{\vec{k^{\prime}}}] |\Delta,\xi\rangle$ will have $\alpha$ such that $\alpha<\alpha_{b+c}$. Thus 
\be
\alpha_a + \alpha < 0, 
\ee
which imply that 
\be
\tilde{O}^{(N)}_{a(b+c)} = 0. 
\ee
We can use similar steps for even level to prove our claim.

\newpage

\appendix

\end{document}